\documentclass[11pt]{article}
\usepackage{authblk}

\usepackage{url}
\usepackage{hyperref}

\usepackage{graphicx}
\usepackage[mathscr]{eucal}
\usepackage{mathrsfs}
\usepackage{amsmath}
\usepackage{placeins}
\usepackage{lineno}
\usepackage{xcolor}
\usepackage[normalem]{ulem}
\usepackage[a4paper,top=3cm,bottom=3cm,left=3.cm,right=3.cm,bindingoffset=0mm]{geometry}

\newcommand{\nm}[1]{\mbox{\tiny #1}}
\newcommand{\nmt}[1]{\mbox{\scalebox{.3}{#1}}}
\newcommand{\ximin}{\xi_{\nm{min}}}
\newcommand{\ximax}{\xi_{\nm{max}}}
\newcommand{\xiA}{\xi_{\nm{A}}}
\newcommand{\xiB}{\xi_{\nm{B}}}

\newcommand{\mx}{m_{\mbox{\tiny X}}}
\newcommand{\yx}{y_{\mbox{\tiny X}}}
\newcommand{\duv}{d_{\mbox{\tiny uv}}}
\newcommand{\duvs}{d_{\mbox{\tiny uv}}^{\mbox{\tiny smear}}}
\newcommand{\duvsm}{d_{\mbox{\tiny uv}}^{\mbox{\tiny sm}}}
\newcommand{\duvg}{d_{\mbox{\tiny uv}}^{\mbox{\tiny gen}}}
\newcommand{\mmu}{\langle\mu\rangle}
\newcommand{\zpv}{z_{\nm{PV}}}
\newcommand{\ztpf}{z_{\nm{ToF}}}
\newcommand{\dz}{\Delta z}
\newcommand{\dzm}{\Delta z_{\nm{min}}}

\begin{document}

\title{Performance studies of Time-of-Flight detectors at LHC}

\author{K. \v Cern\'y}
\affil{\it \small Regional Centre of Advanced Technologies and Materials, Joint Laboratory of Optics of Palack\'y University and Institute of Physics AS CR, Faculty of Science, Palack\'y University, 17.~listopadu~12, 771 46 Olomouc, Czech Republic}
\author{M. Ta\v sevsk\'y}
\affil{\it \small Institute of Physics of the Czech Academy of Sciences, Na Slovance 1999/2, 182~21~Prague~8,~Czech~Republic}
\author{T. S\'ykora}
\author{R. \v Zleb\v c\'ik}
\affil{\it \small Institute of Particle and Nuclear Physics Faculty of Mathematics and Physics, Charles~University, V Hole\v sovi\v ck\'ach 2, 180 00 Prague 8, Czech Republic}
\maketitle

\begin{abstract}
We present results of a toy model study of performance of the Time-of-Flight detectors integrated into forward proton detectors. The goal of the ToF device is 
to suppress effects of additional soft processes (so called pile-up) 
accompanying the hard-scale central diffractive event, characterized by 
two tagged leading protons, one on each side from the interaction point.
The method of mitigation of the pile-up effects exemplified in this study
is based on measuring a difference 
between arrival times of these leading protons at the forward proton 
detectors and hence estimate the z-coordinate of the production vertex. 
We evaluate effects of the pile-up background by studying in detail its components, and 
estimate the performance of the ToF method as a function of
the time and spatial resolution of the ToF device and of the number of pile-up interactions per bunch crossing. 
We also propose a new observable with a potential to efficiently separate central diffractive signal
from the harsh pile-up environment.
\end{abstract}

\section{Introduction}
In diffractive processes, the leading protons produced with high rapidities carry a large fraction of the initial-state beam proton momentum and are separated from the rest of the hadronic final state by the so called large rapidity gap (LRG), i.e. non-exponentially suppressed rapidity interval devoid of particle activity introduced in~\cite{Dokshitzer:1987nc} elaborated in~\cite{Bjorken:1992er}.
Such a behaviour can be described by an exchange of a colorless strong state carrying quantum numbers of vacuum (so-called Pomeron)~\cite{Pomeron}.

\begin{figure}[ht!]
    \centering
    \includegraphics[width=0.19\linewidth,trim={6cm 1cm 9cm 1cm}, clip]{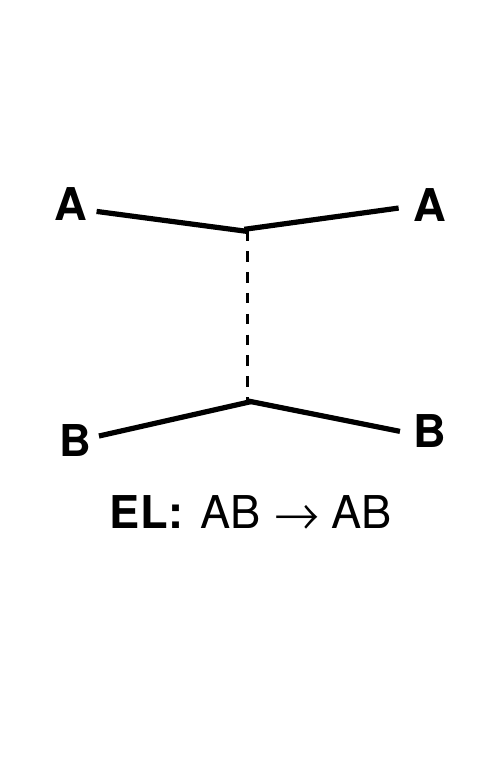}
    \includegraphics[width=0.19\linewidth,trim={6cm 1cm 9cm 1cm}, clip]{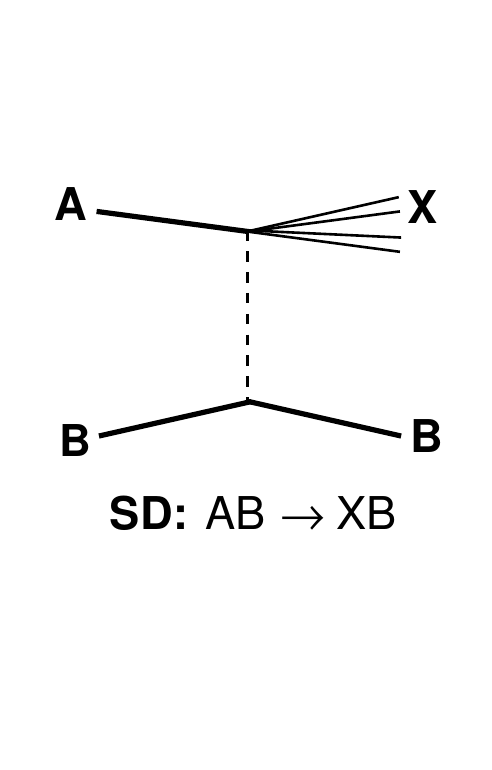}
    \includegraphics[width=0.19\linewidth,trim={6cm 1cm 9cm 1cm}, clip]{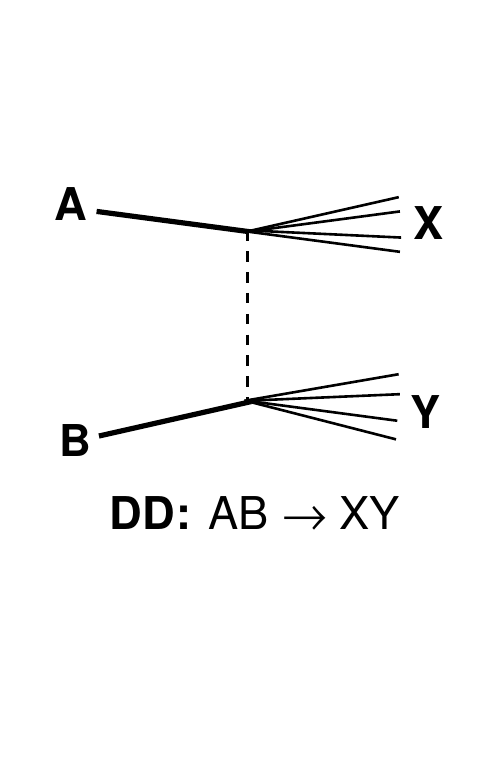}
    \includegraphics[width=0.19\linewidth,trim={6cm 1cm 9cm 1cm}, clip]{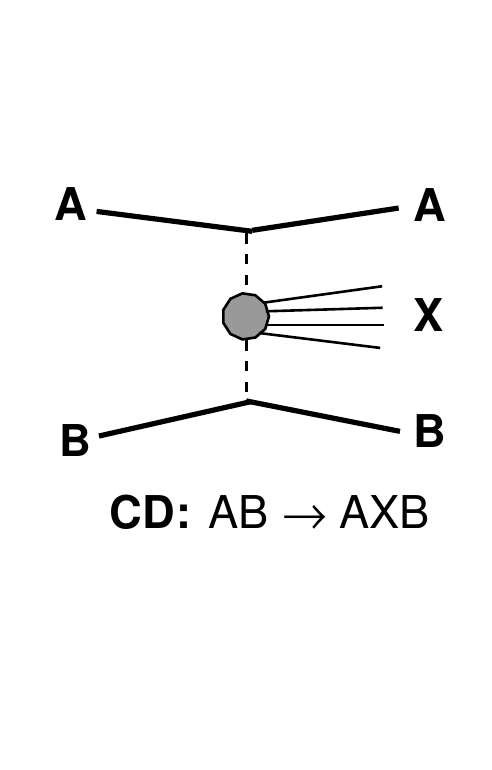}
    \includegraphics[width=0.19\linewidth,trim={6cm 1cm 9cm 1cm}, clip]{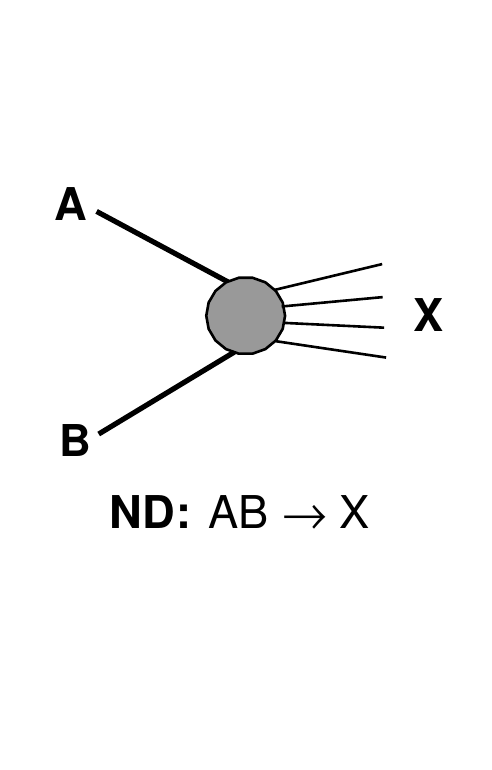}
    \label{fig:SDDDCDEL}
    \caption{The main topologies of processes contributing to the total hadron-hadron cross section; elastic (EL), single-diffractive dissociation (SD), double-diffractive dissociation (DD), central-diffraction (CD) and non-diffractive processes (ND).}
\end{figure}

The diffractive processes in $pp$ collisions at high energies can be divided into several categories according to the topology of the final state, see e.g. ref.~\cite{Zyla:2020zbs}.
We distinguish between the elastic processes (EL), single-diffractive dissociation (SD), double-diffractive dissociation (DD) and central-diffractive processes (CD). Should hard scales be present (represented by large masses or large transverse momenta in the final state) we speak of hard diffractive processes. 
The diagrams in figure~\ref{fig:SDDDCDEL} summarise topologies of the above-defined processes showing also the case of non-diffractive (ND) interactions.

In this text we focus on the CD processes, $pp \rightarrow p X p$, which represents the signal process, while the other topologies represent backgrounds. 
The experimental signature of CD events is characterized by a combination of measurement of the $X$ system in the central detector and the detection of leading protons in the forward proton detectors (FPD) on both sides from the interaction point, referred to as A and B side in the positive and negative direction, respectively, of the interaction axis.
The selection based on such a clear signature becomes less effective if there are more concurrent interactions taking place.
This is typically the case at the LHC~\cite{LHC}, where particle beams organised in particle bunches collide with high instantaneous luminosities.
The mean number of collisions per single bunch crossing, $\mmu$ (also referred to as the pile-up), reached values of about 50 at LHC in 2018~\cite{mu2018} and is expected to be even higher at higher luminosity LHC phases \cite{Azzi:2019yne}.

In this study, we consider two ways the pile-up interactions can mimic the CD signal:
in the first the $X$ system of a non-CD process is reckoned as the signal in the central detector and leading protons from soft diffractive (SD or CD) processes are detected in FPDs, in the other the detected central system $X$ and one of the leading protons come from a genuine CD process, while the other leading proton detected in FPD comes from a soft SD or CD process again.

As shown for the first time in ref.~\cite{firsttofproposal}, the CD signal can effectively be separated from the pile-up effects described above by measuring arrival times of both leading protons to FPDs, equipped by Time-of-Flight (ToF) sub-detectors. In the following, we call this approach the ToF method.
\textcolor{black}{
The ToF detectors can be integrated into FPDs as done in the AFP project~\cite{AFPTDR} in the ATLAS experiment~\cite{ATLAS2008} or CT-PPS project~\cite{CTPPS} in the CMS~\cite{CMS2008} and TOTEM~\cite{TOTEMToF} experiments at the LHC.
}
The difference between the arrival times of the two leading protons produced in the CD interaction (emitted in the opposite directions) is related to the $z$-position of their production vertex in the central detector, $z_{\rm PV}$.
 
It is then evident that by requiring the $z$-positions of the vertex measured by the central detector and by FPDs to match within respective resolutions, the CD event can be separated from the pile-up backgrounds, as documented e.g. in refs.~\cite{excldijets,MarekDM,Tasevsky:2014cpa,RafalDeltaZ,Goncalves:2020saa,STAR}. The efficiency of this separation or the performance of the ToF method depends on number of parameters.

The primary one is the time resolution of the ToF detector.
The other one is the amount of pile-up interactions in the central detector and consequently the number of leading protons produced in one bunch crossing. And the last one is the capability of the ToF detector to distinguish them, i.e. the spatial resolution or granularity of the ToF detector.
Since it is not straightforward to calculate the impact of such effects analytically, we developed a model which we describe in the following.

\section{Model of a single bunch crossing}
The model described in this section simulates the features relevant to the measurement of central-diffractive processes using the ToF detectors in collisions of proton bunches at high instantaneous luminosity, i.e. in presence of pile-up interactions.
\textcolor{black}{
Although the model is used to simulate effects at high $\mmu$ values (up to $\mmu \sim 200$), we do not make use of any additional information from timing detectors in the central detector as planned
for ATLAS in ref.~\cite{HGTDTDR} and CMS in ref.~\cite{CMSMIP} which would naturally lead to improvements in suppressing pile-up backgrounds.
}

\subsection{Basic features of the model}
\textcolor{black}{For each bunch crossing, the model generates a number of vertices according to the Poisson distribution with mean value of $\langle\mu\rangle$. The vertices are generated in the $z$-coordinate (which is the beam collision axis) and time. Both quantities are randomly distributed according to a Gaussian distribution centred at zero and using the width stemming from the typical LHC luminous beamspot width in the $z$-coordinate ($\sigma_{\nm{BS}}$ and $\sigma_{\nm{BS}}/c$ for the spatial and time spreads, respectively). The bunch dimensions in the transverse directions as well as crossing angles of the beams are believed to have negligible impacts on the ToF evaluation and are, therefore, not simulated.}

A special attention is paid to the choice of the primary vertex type. It is assumed that event observables seen by the central detector are reconstructed with respect to the primary vertex. For each event the model generates first one primary vertex of the desired type (CD, ND, SD or DD) and then adds further vertices from pile-up whose types are assigned with a probability proportional to their respective cross sections evaluated using PYTHIA~8~\cite{Pythia8} at $\sqrt{s} = 13$~TeV. Therefore, there are four samples of events generated by the model denoted as CDPV, NDPV, SDPV and DDPV, depending on the process type assigned to the primary vertex.
The EL processes are not expected to contribute, since they do not produce protons capable of reaching the FPDs.

The CD and SD vertices are further complemented by the leading proton(s) whose kinematics are taken from Pythia event files generated beforehand.
The leading protons are subjected to a transport procedure mapping their momentum space kinematics to an auxiliary coordinate space (defined on the A and B sides) by means of which each leading proton is translated to a hit in the ToF detector.
The hits are therefore defined by their local positions and times, where the time is defined by the production vertex time advanced or retarded proportionally to the vertex $z$-position and smeared using a Gaussian function with a width $\sigma_{\nm{t}}$ (ToF detector time resolution) on a random basis.
On top of that an auxiliary detector is assumed to feature a spatial resolution parameter $\sigma_{\nm{x}}$ which can also be reckoned as a granularity parameter and serves to assess the impact of multi-hit events in the detector. 

The initial setting of the beamspot size (50~mm), the size (2~cm) and location (1~mm from the nominal beam) of the detector as well as of the transport mapping (approximately linear in $\xi$) are chosen such that they do not depart too much from the realistic values available at the LHC.
The main degrees of freedom of the model can thus be summarised as being represented by the values of $\langle\mu\rangle$, $\sigma_{\mbox{\tiny BS}}$, $\sigma_{\mbox{\tiny t}}$ and $\sigma_{\mbox{\tiny x}}$.
An additional freedom of the model is also introduced by the choice of the ToF detector positions and the transport details as well as by the methods used for dealing with multi-hit signals in the ToF detectors. \textcolor{black}{It is also assumed that each arm is synchronized with the reference clock at a level of $2$~ps (see Section $4.6.5.2$ of ref.~\cite{AFPTDR})}.

Due to conservation laws the kinematics of the leading protons and the centrally produced system $X$ of the primary CD process ($pp \rightarrow pXp$) are correlated.
\textcolor{black}{
Strong correlations (even perfect within detector resolutions) are expected for exclusive processes, where the energy losses of the leading protons are fully transferred to the system $X$. In the case of inclusive CD processes, proton energy losses are shared between the system $X$ and Pomeron remnants which usually continue in the direction of the incoming proton thus leave the central detector partly undetected. This leads to correlations which are significantly worse than in the case of exclusive processes.} 
The correlations (or the lack of them) also allow for a better (or worse) separation of the genuine CD process from the non-CD ones. Because there are various processes that can be studied each with the specific experimental signature (topology of the $X$), dedicated cuts must always be applied to suppress various backgrounds including effects of pile-up. Any detailed analysis of the $X$ topology and of the optimal cuts goes beyond the scope of the presented toy model. The model studies the kinematics and time information of the leading protons only.

\subsection{Kinematics of signal events}
The leading protons produced in the diffractive processes can be described in terms of the relative momentum loss ($\xi$), the Mandelstam ($t$) variable and the azimuthal angle ($\phi$) (not considered here, usually integrated over), defined as

\begin{eqnarray}
\xi = \frac{E_{\mbox{\tiny beam}} - E_{\mbox{\tiny p}}}{E_{\mbox{\tiny beam}}},
\label{eq:mx} \\
t = (P_{\nm{beam}}-P_{\nm{p}})^{2},
\label{eq:yx}
\end{eqnarray}

where $E_{\mbox{\tiny p}}$ ($E_{\mbox{\tiny beam}}$) represents the energy component of the leading (beam) proton four-vector, i.e. $P_{\nm{beam}}$ ($P_{\nm{p}}$). The role of the $t$ variable is negligible in the model. The most important is the energy of the beam proton available for the central interaction, i.e. $\xi E_{\nm{beam}}$.

In the case of the $pp \rightarrow pXp$ interactions the invariant mass and rapidity of the $X$ system can be unambiguously related to the $\xi$ fractions of the two leading protons as

\begin{eqnarray}
m_{\nm{X}} = \sqrt{s \xi_{A} \xi_{B}}&\mbox{    and    }& y_{\nm{X}} = \frac{1}{2} \mbox{ln}\left(\frac{\xi_{A}}{\xi_{B}}\right), 
\label{eq:mxyx} \\
\xiA = \frac{m_{\nm{X}}\,e^{y_{\nm{X}}}}{\sqrt{s}}&\mbox{    and    }&\xiB = \frac{m_{\nm{X}}\,e^{-y_{\nm{X}}}}{\sqrt{s}}.
\label{eq:xiAxiB}
\end{eqnarray}

\textcolor{black}{
Of course, the above relations hold in case the $X$ system is well resolved in the central detector which favours the exclusive processes in practice.
}

The $m_{\nm{X}}$ and $y_{\nm{X}}$ observables define the kinematic plane of CD processes. For a fixed value of one of the $\xi$ fractions the $y_{\nm{X}}$ variable is a linear function of logarithm of $m_{\nm{X}}$ which simplifies the interpretation of the ($m_{\nm{X}}, y_{\nm{X}}$) kinematic plane. The range of kinematics where both leading protons end up in the acceptance of both ToF detectors, the so-called double-tag (DTAG) range, is very well defined then. A single variable (called $d_{\nm{uv}}$ here) can be constructed to assess the proximity of the CD event kinematics to the DTAG range, where the DTAG range is governed by the position and size of the ToF detectors. The definition of the $d_{\nm{uv}}$ discriminator is given in appendix~\ref{app:dbltagdiscriminator}.

In figure~\ref{fig:genkin}a) the generated kinematics ($m_{\nm{X}}$, $y_{\nm{X}}$) of the primary CD processes in the CDPV sample is plotted for events with $\langle\mu\rangle = 10$ tagged on both sides (in detectors of $2$~cm size placed $1$~mm from the nominal position discussed with the transport in the next section). The DTAG range (indicated by the blue line) is visibly well enhanced as the intersection of two strips that correspond to events with one leading proton tagged (single-tag, STAG). 

The effect of pile-up interactions is visualized by shaded areas outside the STAG and DTAG areas.

As mentioned above, kinematics of the final state $X$ system are not primarily analysed in the model. However, for the CD processes it makes sense to assume that the reconstruction of $X$ would lead to values of $\mx$ and $\yx$ smeared by experimental resolutions of the central detector around the generated values which can be obtained via leading proton kinematics. Conservative resolutions of $30 \%$ and $0.3$ for the reconstruction of $\mx$ and $\yx$, respectively, are propagated to the $\duv$ calculation, thus corresponding to $\duvs$ (or $\duvsm$ for brevity). The functionality of the $\duv$ selection is evidenced by the b) and c) panels of figure~\ref{fig:genkin}. 

In figure~\ref{fig:genkin}b) we document that the kinematics of the CDPV sample events, selected by the cut $\duvs = 1$, are constrained to a proximity of the DTAG range.
Complementarily, in figure~\ref{fig:genkin}c) it is shown how relatively little of the generated kinematics leak outside the DTAG range due to smearing, if a $\duvg = 1$ selection is applied. 

Eventually, the distribution of the $\duvs$ variable and its $\duvg$ components are shown in figure~\ref{fig:genkin}d). The $\duv$ discriminator clearly has a potential to suppress the contribution of CD events generated outside the DTAG acceptance which are falsely tagged due to the contribution of leading protons from pile-up. It represents an alternative approach to cuts realised as simple matching cuts between the $\mx$ and $\yx$ quantities measured in the central detector and the FPDs. The adequate value of the $\duvs$ cut would be a matter of optimisation (not done here) depending on the actual precision of the $\mx$ and $\yx$ measurements.

\begin{figure}
\begin{center}
\includegraphics[width=0.95\linewidth, trim={0cm 0cm 0cm 0cm},clip]{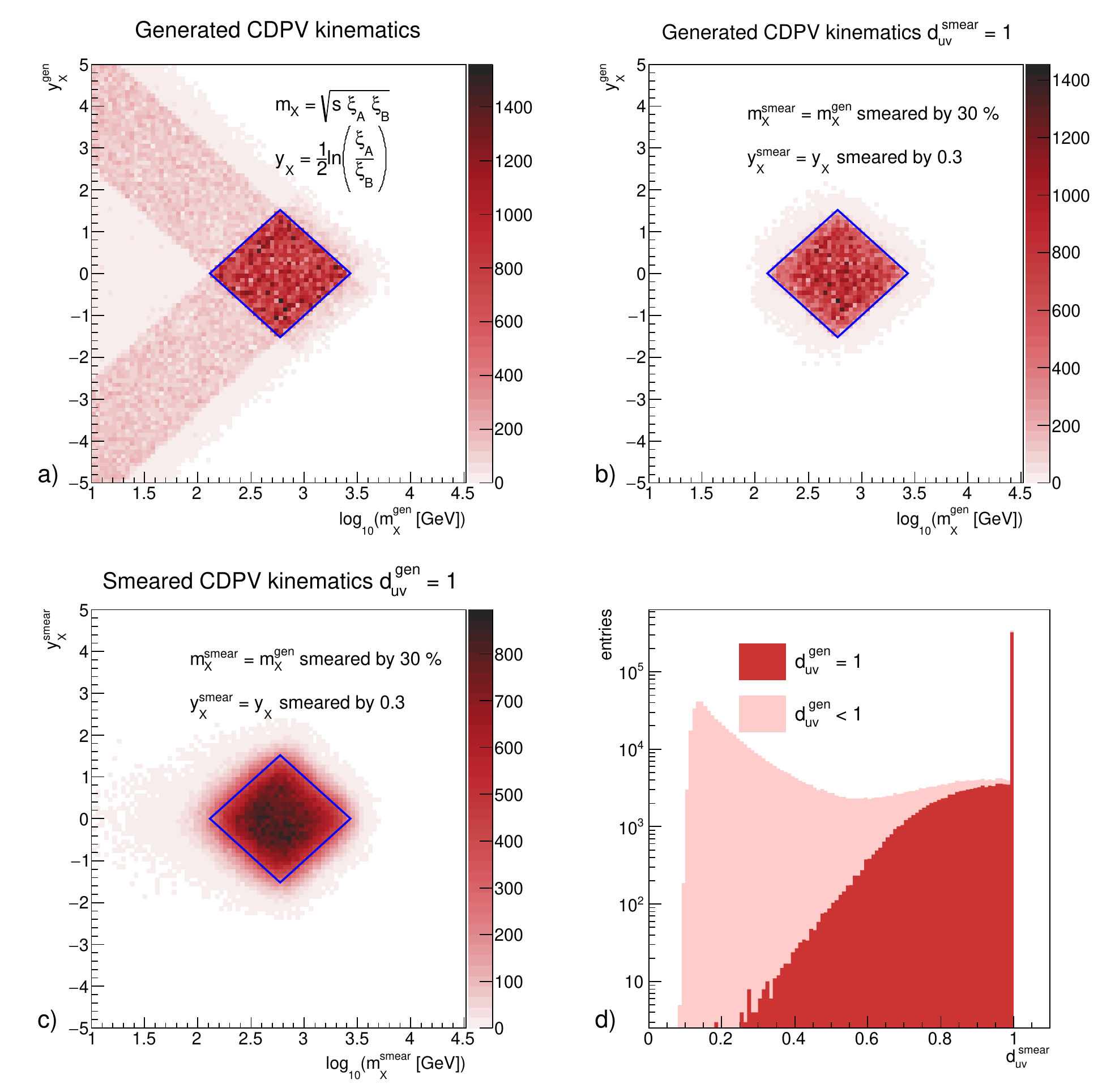}
\end{center}
\caption{Generated and smeared kinematics of central diffractive processes in the CDPV sample overlayed with pile-up ($\langle\mu\rangle = 10$) with signal in the detectors of $2$~cm size located $1$~mm from the nominal beam. The smearing of generated values is done using resolutions of 30\% for the $\mx$ and 0.3 for the $\yx$ variable. 
a) Generated CD kinematics together with the DTAG range indicated by the blue line, b) effect of the $\duvs$ cut on generated kinematics, c) smeared kinematics selected by the $\duvg$ cut, d) distribution of the $\duvs$ discriminant obtained from the smeared generated kinematics. The contribution of CD events with true kinematics generated in the DTAG range is emphasised by the dark red histogram, the events outside the true DTAG range are represented by the pale red histogram.}
\label{fig:genkin}
\end{figure}

\subsection{The leading proton transport}
The leading proton kinematics given in terms of $\xi$, $t$ and $\phi$ affects the probability of detection of leading protons in FPDs. We use a simple transport linear in $\xi$ disregarding the role of $t$ and $\phi$, implemented analytically as a $\xi \rightarrow x$ mapping defined as follows

\begin{equation}
x = 100\, \xi \left[\mbox{mm}\right],
\label{eq:transport}
\end{equation}
where the $x$ value represents a hit position measured in the ToF detector.

The detectors are defined as sensitive volumes measuring the $x$-coordinate in an auxiliary $x$-space, where the $x = 0$ point represents the nominal beam. The detectors are given a length of $20$~mm with the detector edge placed at $x = 1$~mm from the origin of coordinates as a baseline position, which limits the measurable $\xi$ values to the range of $0.01$--$0.21$. 

The detector dimensions, position and the resulting range of accessible $\xi$ values are similar to those usually achievable by FPDs for hard diffractive physics at LHC experiments. Possible non-linearities and smearings of the mapping (due to the limited position resolution, beamspot smearing and folding of the $t$ and $\phi$ kinematics) are neglected here. The mapping is for example useful to study the impact of limited granularity which is realised here as an equidistant division of the sensitive detector range to cells of the $\sigma_{x}$ size.

\subsection{Analysis of generated events}
\label{sec:analysisofgeneratedevents}
The generated event data contain information about the primary vertex $z$-position, $z_{\nm{PV}}$, and positions and times of hits measured in the two FPDs. A hit filtering procedure is adopted such that hits occupying the same detector cell are merged into one new hit with a time stamp of the earliest one.
 
The case of $\sigma_{\nm{x}} = 0$ is a special case of a detector with an ideal granularity, i.e. no hit filtering.

For each particular pair of hits from opposite sides, the $z_{\nm{ToF}}$ variable can be reconstructed as
\begin{equation}
z_{\mbox{\tiny ToF}} = -\frac{c}{2} (t_{\mbox{\tiny A}} - t_{\mbox{\tiny B}}),
\label{eq:ztofformula}
\end{equation}  

We assume that FPDs at the A and the B sides are located at the same distances from the interaction point. 
If there are multiple pairs of hits, a list of $z_{\nm{ToF}}$ vertices is obtained.

On the event-by-event basis, each of the list of $z_{\mbox{\tiny ToF}}$ vertices is compared with the value of $z_{\nm{PV}}$ thereby forming a $\Delta z = z_{\nm{PV}} - z_{\mbox{\tiny ToF}}$ distribution.
While the width of the $\Delta z$ distribution in the signal CDPV sample is proportional to $c\sigma_t$, the widths in the background samples are much broader and depend on $\sigma_{\rm BS}$.
More precisely, the width of the signal $\Delta{z}$ distributions neglecting the primary vertex reconstruction uncertainty is given by $c\sigma_{t}/\sqrt{2}$ which is used as the cut applied to select the genuine CD events (so called $\Delta z$ veto), i.e.
\begin{equation}
\left|\Delta{z}\right| < \frac{c}{\sqrt{2}}\sigma_{t}.
\label{eq:deltazcut}
\end{equation}  

There are two kinds of background contributions to the $\Delta{z}$ distribution. The first one originates from independent combinations of the ($z_{\nm{ToF}}$, $z_{\nm{PV}}$) values, where $z_{\nm{ToF}}$ is reconstructed from hits generated by pile-up interactions only.
A single Gaussian shape of such contribution is expected with a width of $\sqrt{2} \sigma_{\mbox{\tiny BS}}$~\footnote{if $\sigma_{t}$ is neglected}. The second kind of background (partially tagged) can be expected to contribute in the CDPV sample only, where one of the hits used for the $z_{\nm{ToF}}$ calculation was caused by a genuine leading proton generated in the CD event in the primary vertex. The width of this partially tagged background is equal to $\sigma_{\nm{BS}}$. The $\Delta z$ widths of all considered backgrounds are discussed in detail in the appendix~\ref{app:widths}.

\begin{figure}[ttt]
\begin{center}
\includegraphics[width=0.37\linewidth, trim={8.5cm 3.7cm 7.1cm 4.7cm},clip]{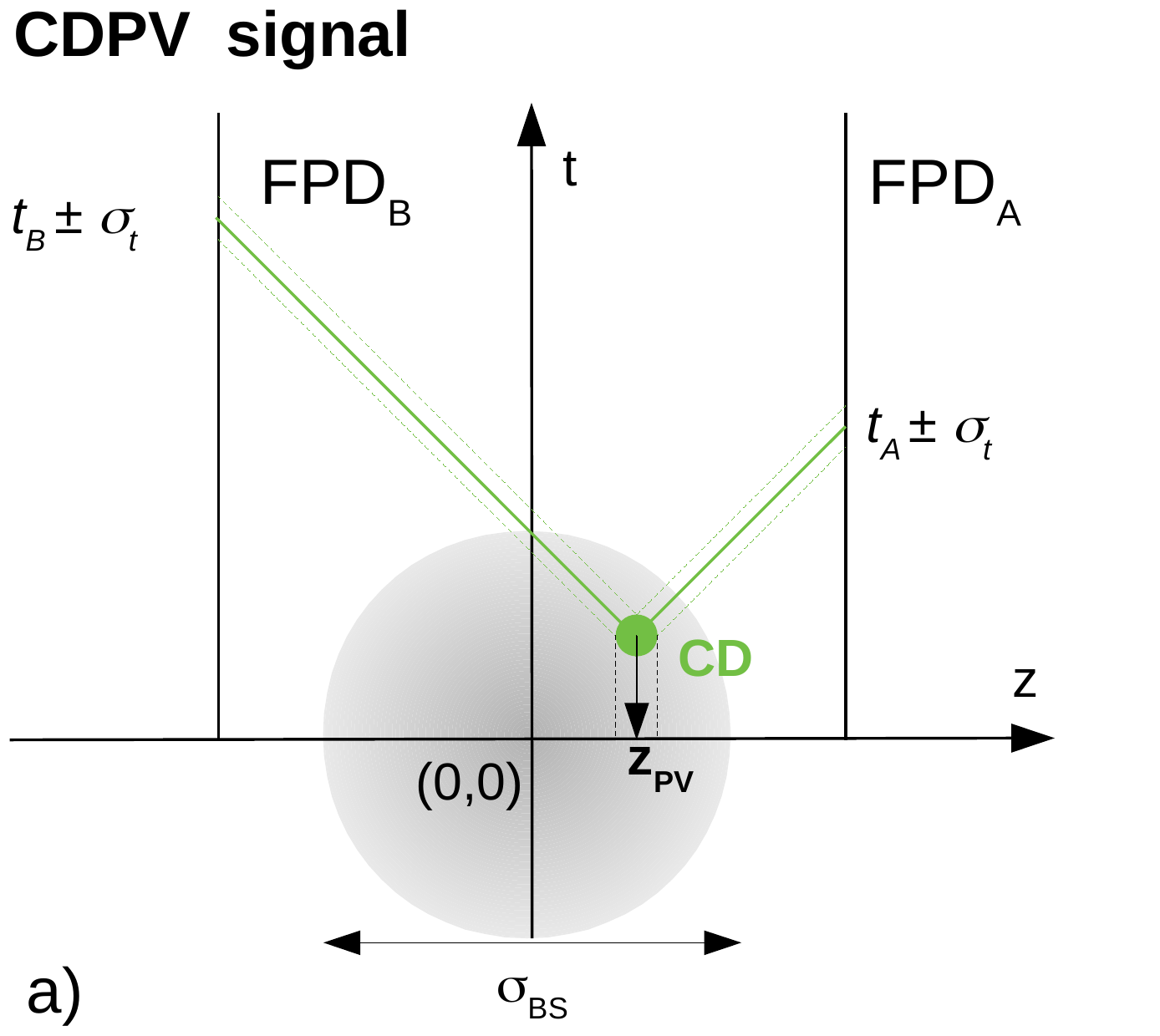}
\hspace{1cm}
\includegraphics[width=0.37\linewidth, trim={8.5cm 3.7cm 7.1cm 4.7cm},clip]{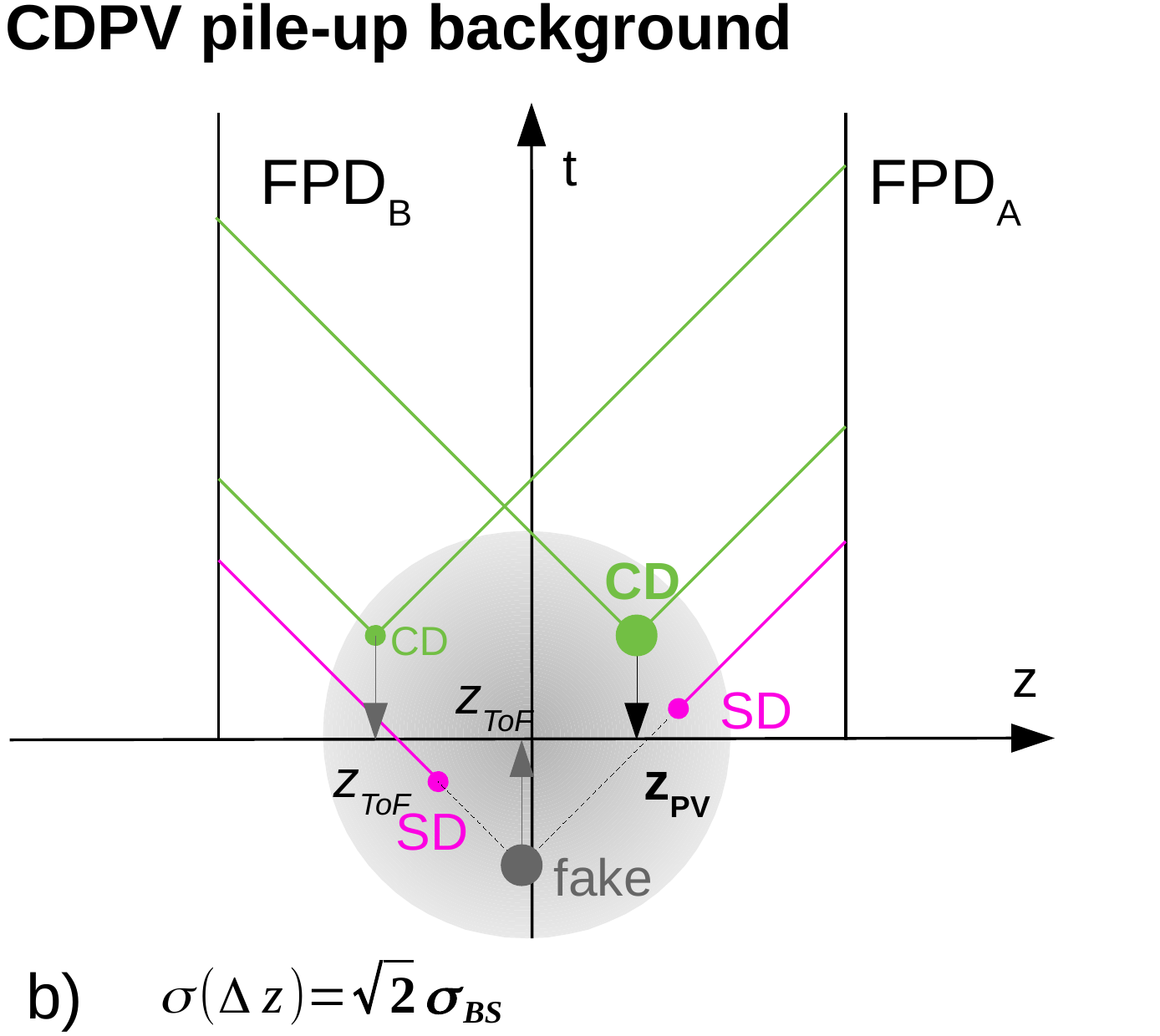}
\\
\vspace{0.5cm}
\includegraphics[width=0.37\linewidth, trim={8.5cm 3.7cm 7.1cm 4.7cm},clip]{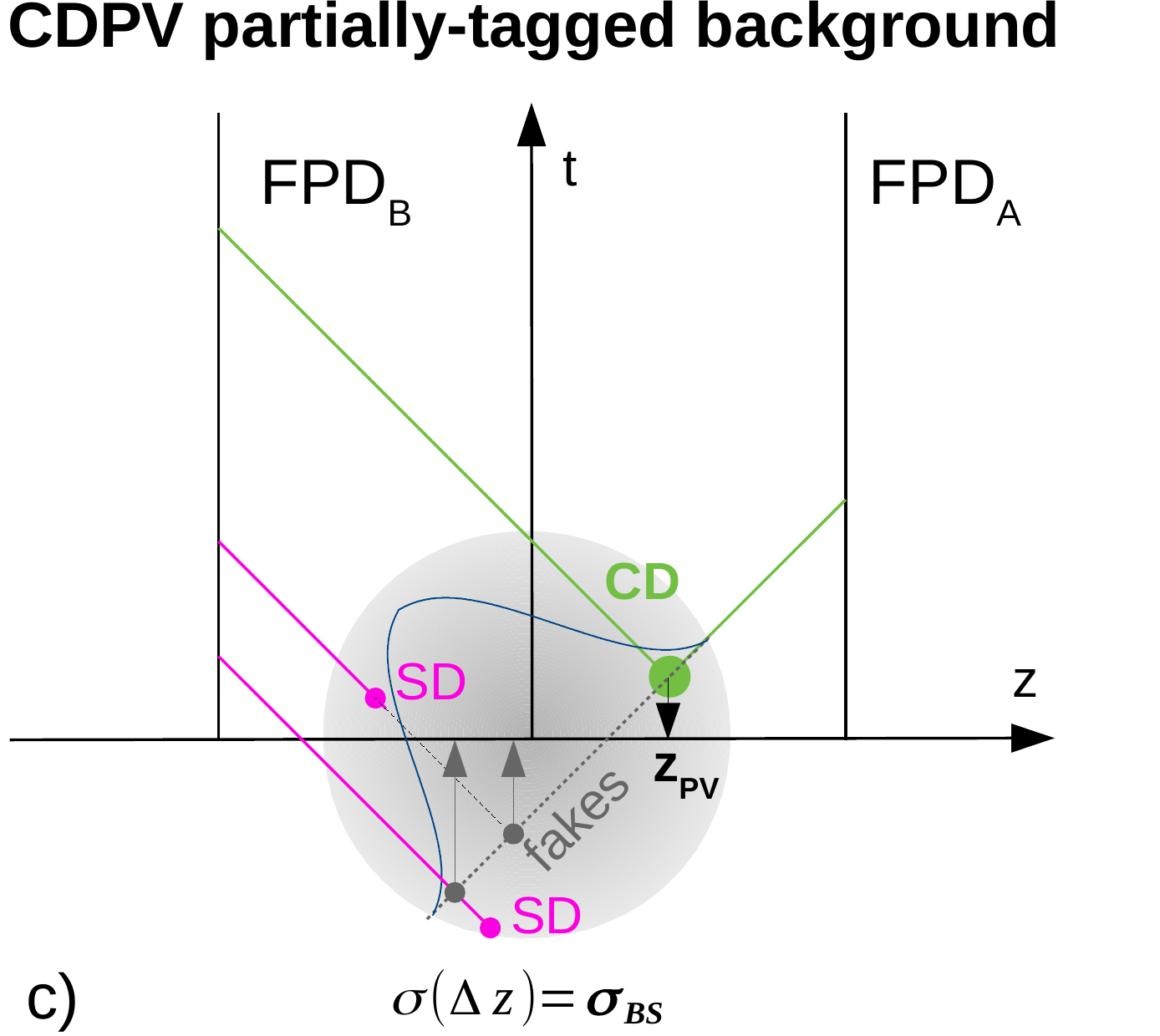}
\hspace{1cm}
\includegraphics[width=0.37 \linewidth, trim={8.5cm 3.7cm 7.1cm 4.7cm},clip]{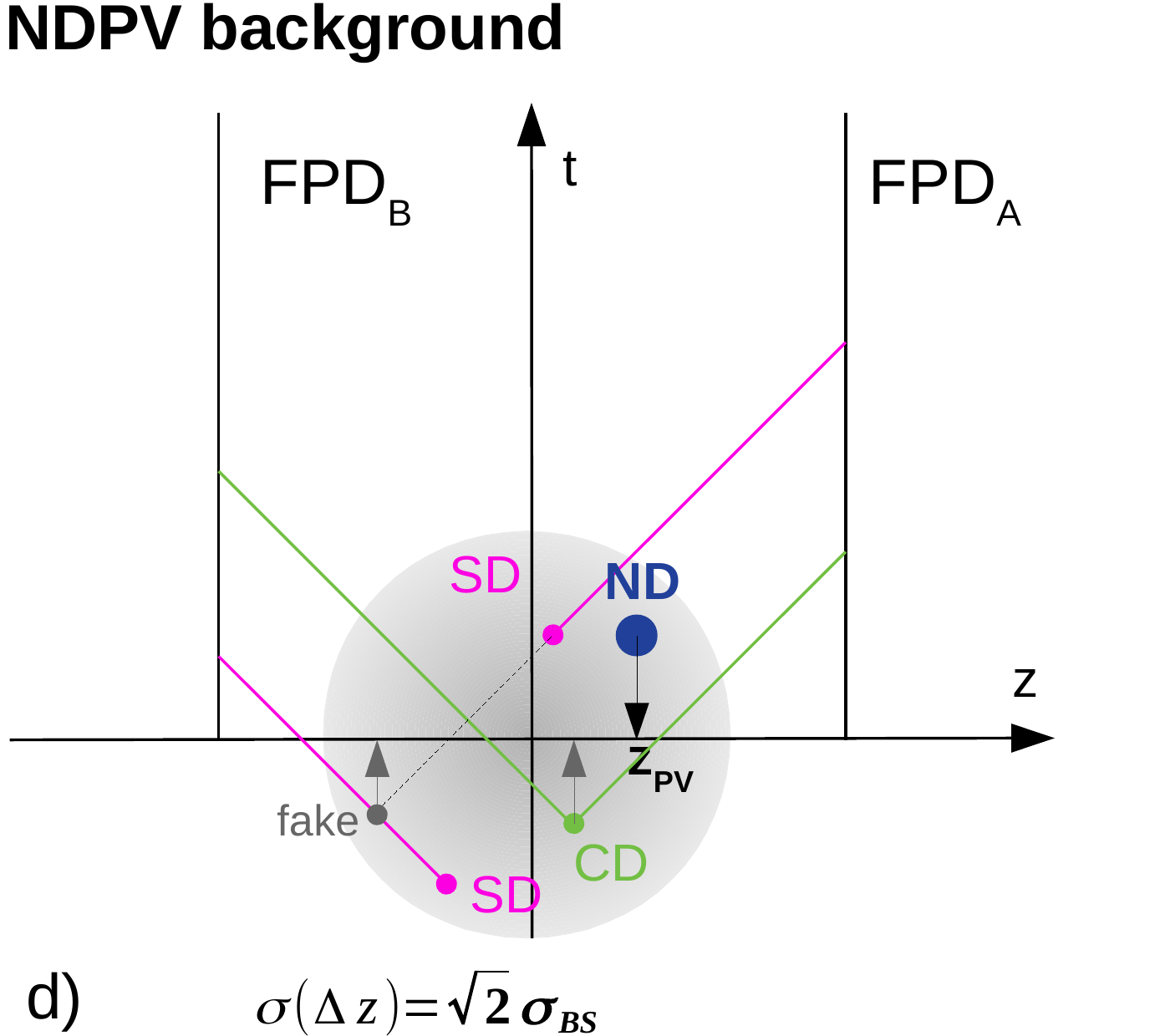}
\end{center}
\caption{The space-time diagrams describing the rationale behind the use of Time-of-Flight method. 
The horizontal and vertical axes correspond to the z-coordinate and time, respectively, where the centre of the beam spot (represented by the large shaded circle) is located in the origin of coordinates. The positions of the time-measuring FPD detectors are indicated by the vertical lines at fixed z positions. The world lines of the leading protons travelling at speed of light are sketched with $\pm45^{\circ}$ lines and primary vertices are represented by circles (green for CD, magenta for SD and blue for ND processes) and marked by $\zpv$. The vertices coming from ToF measurements and marked by $\ztpf$ (represented by small grey circles) are caused by a fake double-tag of two leading protons from two unrelated events and are put at the intersection of corresponding lines going in opposite directions.
}
\label{fig:tofevents}
\end{figure}

It is convenient to use space-time coordinates to depict the rationale behind the ToF method as demonstrated in figure~\ref{fig:tofevents}, 
where topologies of signal and different backgrounds are shown
(the space-time coordinates are scaled to equal display units). The Gaussian bunch-crossing contours of width $\sigma_{\nm{BS}}$ are indicated by the shaded circle. The leading protons are indicated by $\pm45^{\circ}$ lines reaching positions of FPD detectors on sides A and B. In figure~\ref{fig:tofevents}a) the simplest CD process from the CDPV sample is visualized where both detectors provide information consistent with the primary vertex position within the $\sigma_{t}$ range shown by dashed lines along the leading proton nominal lines. In figure~\ref{fig:tofevents}b) the pile-up background in the CDPV cases is sketched, which produces the identical background shape as the non-CDPV samples, i.e. NDPV in figure~\ref{fig:tofevents}d) for instance. This is caused by the fact that the fake (having in mind the spurious vertices reconstructed from SD events) and possibly also the non-primary CD vertices are distributed independently of $z_{\nm{PV}}$, both with $\sigma_{\nm{BS}}$ widths. The origin of the partially tagged CDPV background is also shown in figure~\ref{fig:tofevents}c) where the main ingredient is the fact that one of the measured proton arrival times is coming from the actual CD-primary vertex. In order to form pairs, pairs of hits from opposite sides are formed which leads to reconstruction of fake vertices that are no longer distributed independently in space and time, they populate a (z,t) world-line defined by the tagged proton from the primary CD process.

\begin{figure}[t]
\begin{center}
\includegraphics[width=0.93\linewidth, trim={1cm 0.3cm 1.1cm 0.3cm},clip]{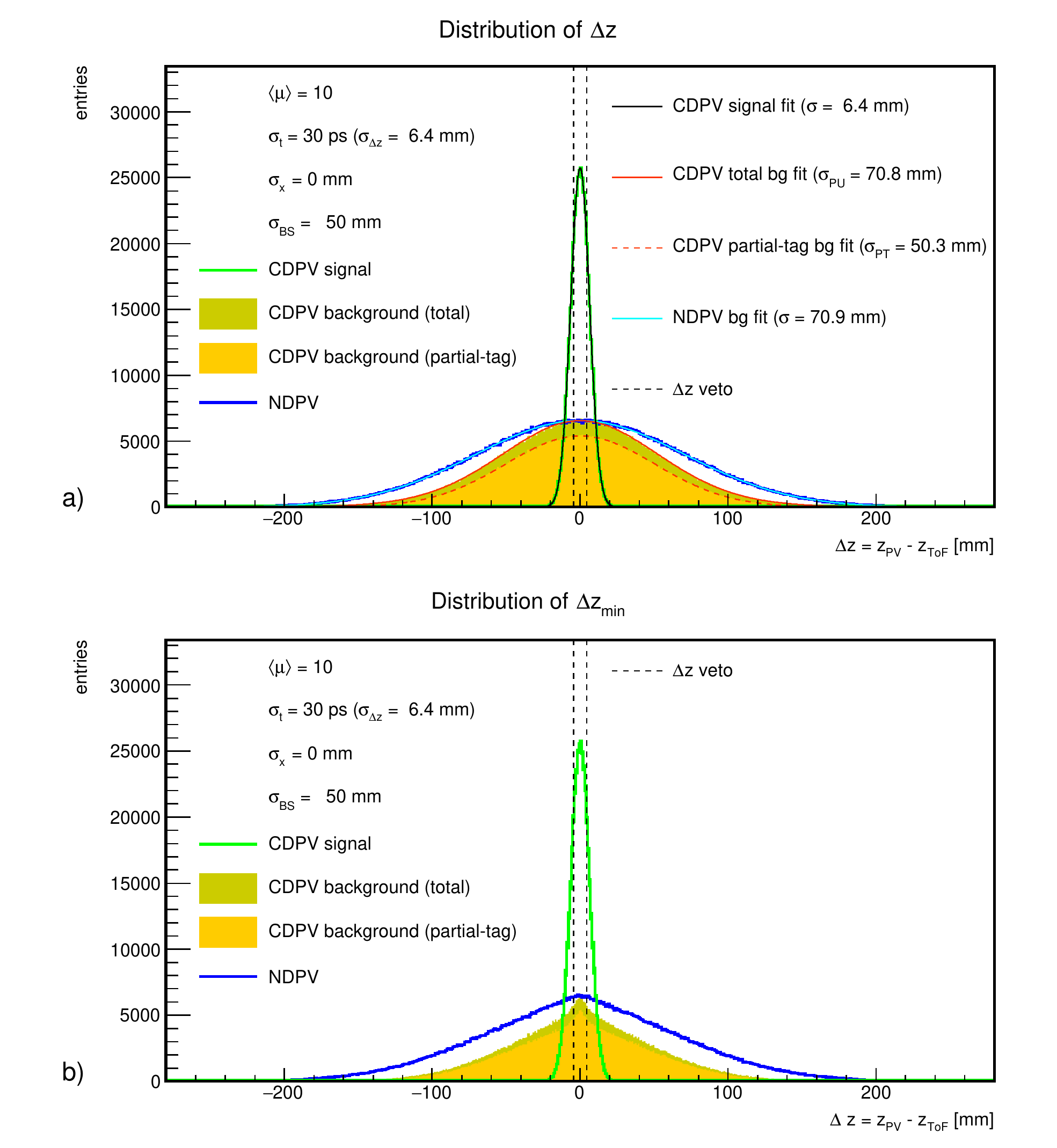}
\caption{Unnormalized $\Delta{z}$ distributions in the CDPV and NDPV samples generated using $\mmu$ = 10, $\sigma_{t} = 30$~ps, $\sigma_{x} = 0$ and $z_{\nm{BS}} = 50$~mm. a) The signal is shown by the light green histogram and total CDPV, partially-tagged CDPV and NDPV backgrounds by the dark yellow, yellow and blue histograms, respectively. Fits to the distributions are shown by lines. The fitted widths of the CDPV background components are denoted as $\sigma_{\nm{PU}}$ and $\sigma_{\nm{PT}}$ for the pile-up and partial-tag contributions. The $\Delta z$ veto is indicated by the vertical dashed black lines. b) The same distributions are shown for the $\Delta z_{\nm{min}}$ method with the same color coding as in a).
}
\label{fig:deltaz}
\end{center}
\end{figure}

The points discussed above are illustrated in the plot in figure~\ref{fig:deltaz}a), where the $\Delta{z}$ distributions are plotted for CDPV and NDPV samples for $\mmu$ = 10, $\sigma_{t} = 30$~ps and $\sigma_{x} = 0$. The signal and background contributions to the total $\dz$ distribution in the CDPV sample are shown where the fractional partial-tag CDPV background is indicated and fitted separately. The fitted widths of the signal (6.4~mm) and background distributions (50.3~mm and 70.8~mm~\footnote{where $50\sqrt{2}\simeq 70.7$}) are consistent with the input values of $\sigma_{t}$ = 30~ps and $\sigma_{\nm{BS}}$ = 50~mm. In the NDPV sample the $\dz$ variable contains only one background component described by a single Gaussian fit of a 70.9~mm width as expected. In figure~\ref{fig:deltaz}b) results for the same components to the total $\Delta z$ distribution are shown for an alternative definition of the $\dz$ variable, namely only a single $\dz$ value closest to the $\zpv$ is considered per event, denoted as $\dzm$. The shapes of the  signal and background components are not Gaussian and no attempt to perform fits was made. The advantage of this method is that one has just one value of $\Delta z$ = $\dzm$ per event to deal with.

\FloatBarrier

\section{Results}
In this section we discuss results obtained from the toy model and the performance of the ToF method to extract the CD signal from pile-up backgrounds only on the basis of kinematics of forward protons detected in ToF detectors. 
We assume that the primary selection of a CD process under study in an offline analysis 
would be based on a central-detector trigger part followed by a proper analysis of the hadronic final state $X$. 

In the following analysis the events with ND and SD processes in the primary vertex are assumed to represent backgrounds and to contaminate the sample of selected events. The contribution of the DD background ($pp \rightarrow XY$) process is neglected thanks to a very different final state comprising two forward-going systems $X$ and $Y$, separated by a large rapidity gap, and missing leading protons produced directly. The EL processes are naturally neglected as well because they do not mimic neither the central system $X$, nor the leading protons.

The probability of the signal observation in the two ToF detectors in a single event represents a first observable measured by the model. It is defined as

\begin{equation}
P_{\nm{DTAG}} = \frac{ N_{\nm{DTAG}}} {N_{\nm{gen}}},
\end{equation}

where $N_{\nm{gen}}$ is the number of events needed to obtain $N_{\nm{DTAG}}$ events that satisfy the DTAG condition.

For the latter then the probability that a given event is selected by having at least one entry in the $\Delta z$ window, i.e. passing the $\dz$ veto (eq.~\eqref{eq:deltazcut}), is calculated as
\begin{equation}
P_{\Delta z} = \frac{N_{\Delta z\nm{-cut}}}{N_{\nm{DTAG}}},
\label{eq:PDELTAZ}
\end{equation}
where $N_{\Delta z\nm{-cut}}$ denotes the number of events with at least one entry in the $\dz$ window , i.e. involving signal or pile-up protons. 
Finally, the probability that the $\dz$ veto is satisfied through detection of the primary CD process is denoted by $P_{\Delta z}^{\nm{signal}}$. The $P_{\Delta z}^{\nm{signal}}$ quantity is therefore only defined on the CD signal events from the CDPV sample, i.e. those having two genuine leading protons originating in the primary vertex and not from accompanying pile-up events.

The $P_{\nm{DTAG}}$ and $P_{\Delta z}$ are also calculated for the case when the events of the CDPV sample are preselected by the $\duv$ cut. Since this cut selects events with kinematics close to the DTAG range, $P_{\nm{DTAG}}$ and $P_{\Delta z}$ values are naturally enhanced.

\begin{figure}[ttt]
\begin{center}
\includegraphics[width=0.47\linewidth, trim={0.7cm 0cm 0.7cm 0cm},clip]{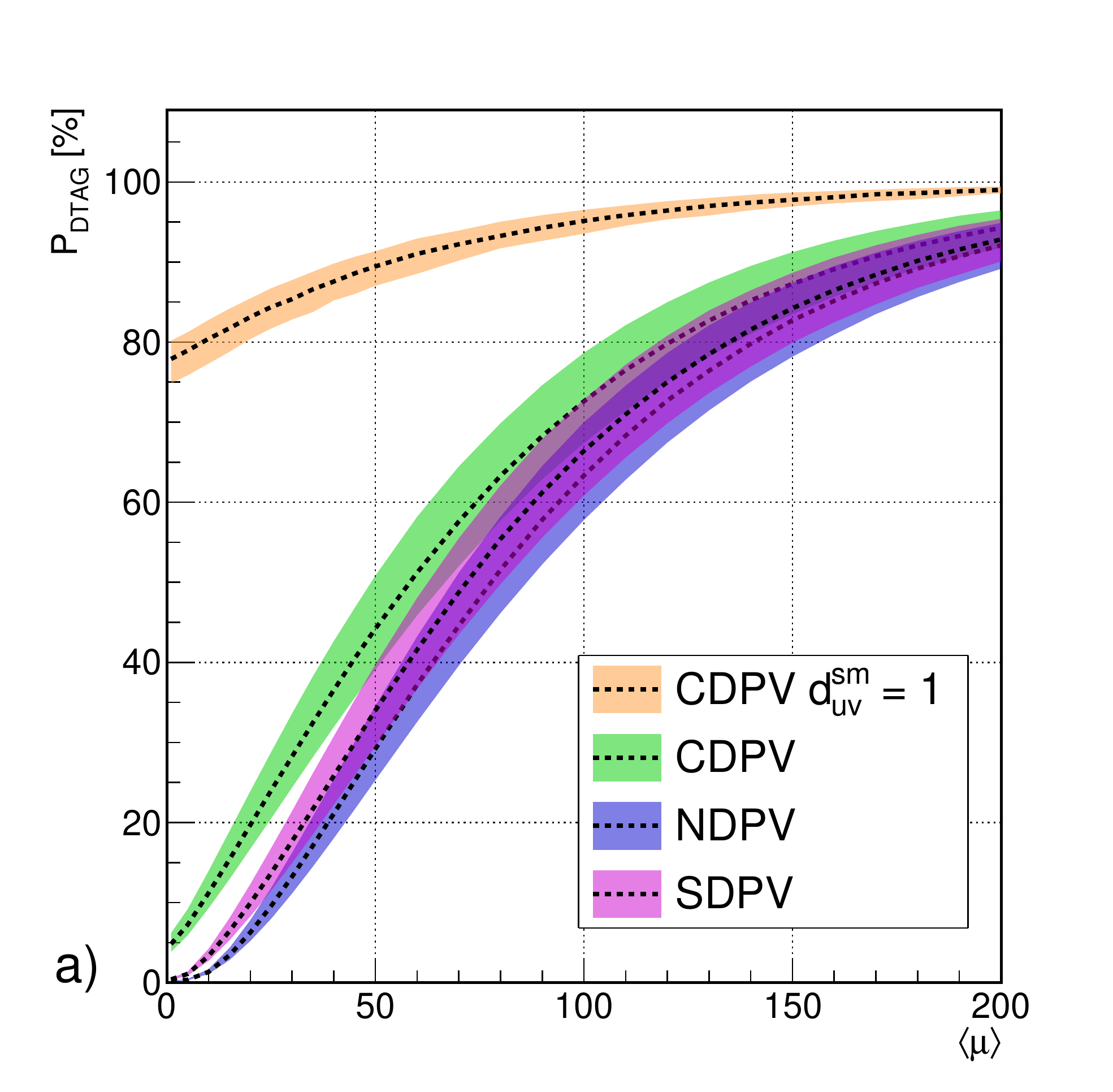}
\includegraphics[width=0.47\linewidth, trim={0.7cm 0cm 0.7cm 0cm},clip]{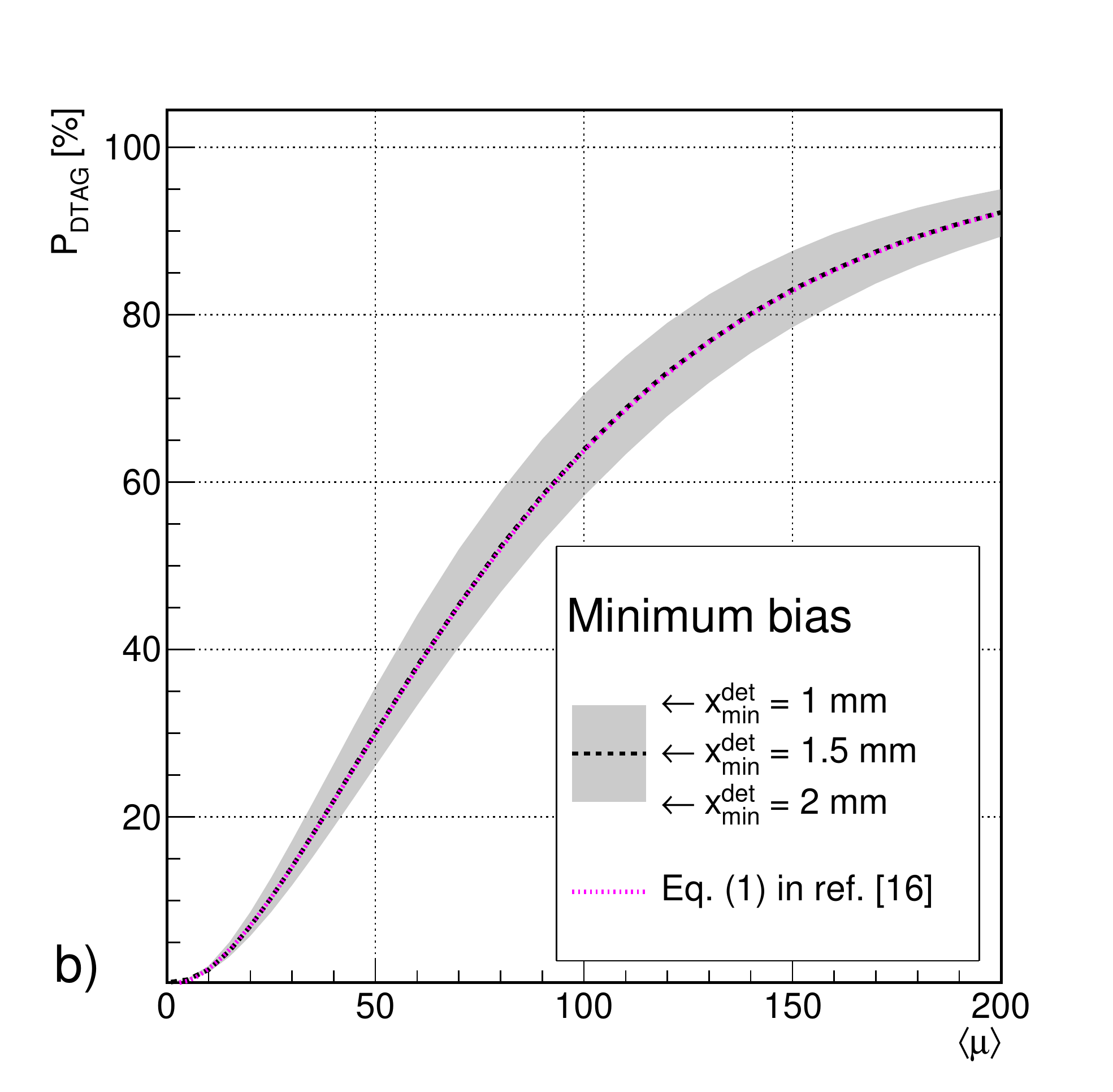}
\caption{a) ToF double-tag probability as a function of $\mmu$ for three detector edge positions shown as coloured bands where values from the top (bottom) of the bands correspond to $x_{\nm{min}}^{\nm{det}} = 1~(2)$~mm and the middle dashed lines indicate the $x_{\nm{min}}^{\nm{det}}= 1.5$~mm results. The green, magenta and blue bands represent results from the CDPV, SDPV and NDPV samples, respectively, the orange band shows results from the CDPV sample where the generated kinematics are constrained by $\duvs = 1$. b) $P_{\mbox{\nm{DTAG}}}$ values obtained as weighted average of the CDPV, NDPV and SDPV values and represented by the white dashed line are compared with the prediction based on eq.~($1$) in ref.~\cite{Tasevsky:2014cpa} shown by the magenta dotted line for the case of $x_{\nm{min}}^{\nm{det}} = 1.5$~mm.
}
\label{fig:PDTAG}
\end{center}
\begin{center}
\includegraphics[width=0.45\linewidth, trim={0.5cm 0.2cm 0.5cm 0.5cm},clip]{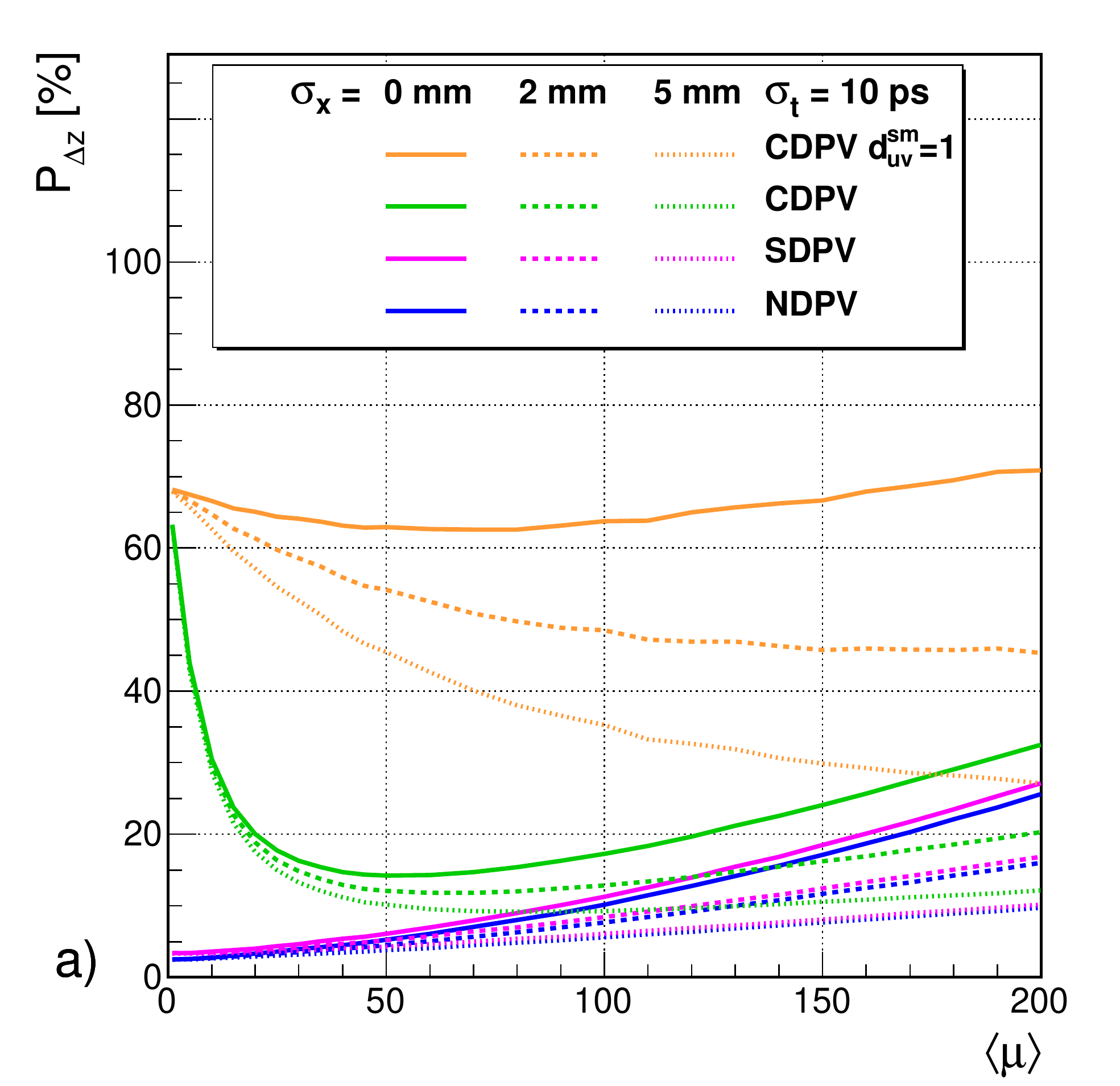}
\includegraphics[width=0.45\linewidth, trim={0.5cm 0.2cm 0.5cm 0.5cm},clip]{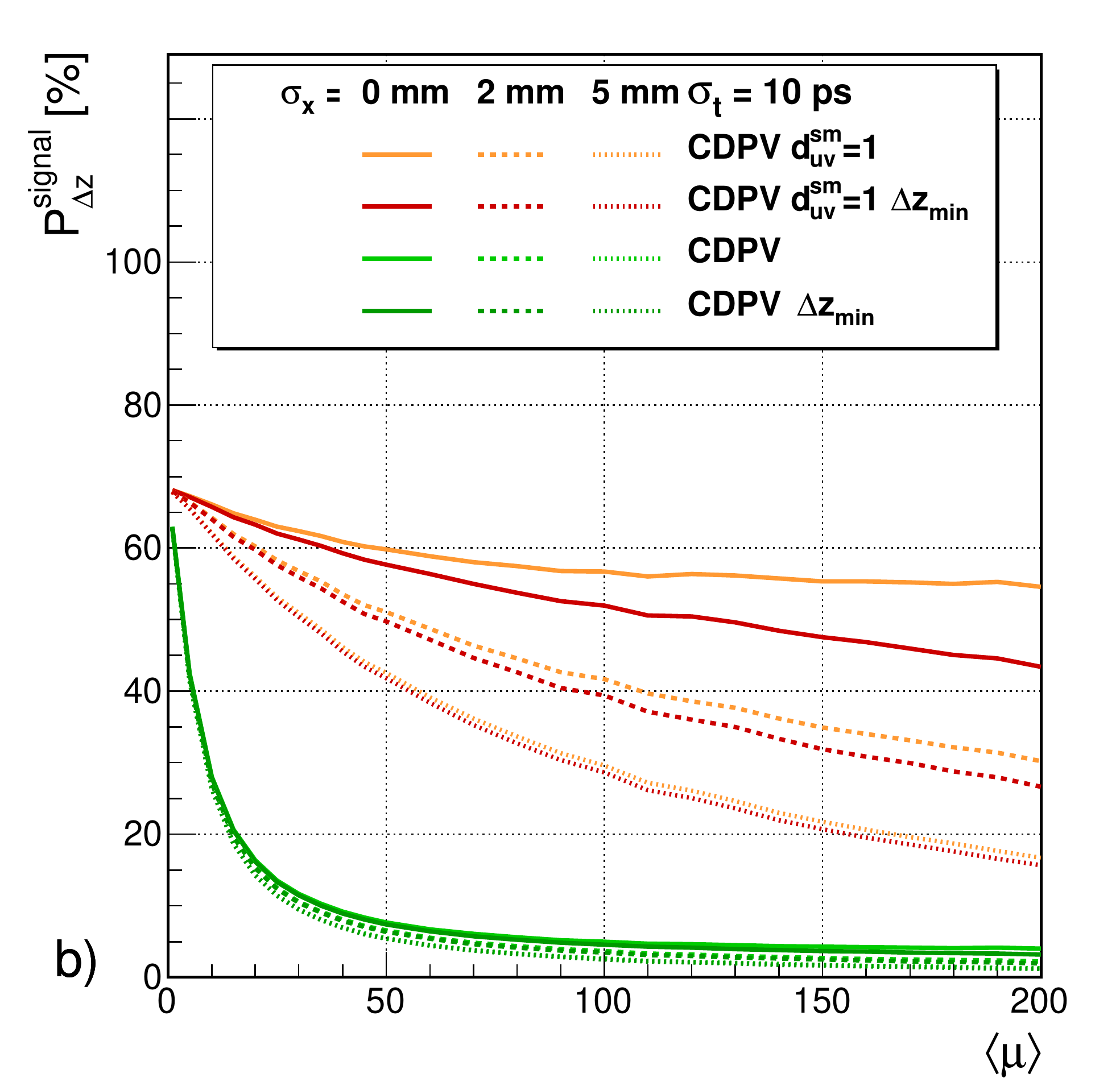}
\caption{
$P_{\Delta z}$ and $P_{\Delta z}^{\nm{signal}}$ probabilities as a function of $\mmu$ for $\sigma_t = 10$~ps and three $\sigma_{x}$ values of 0, 1 and 2~mm shown by solid, dashed and dotted lines. a) $P_{\Delta z}$ of the CDPV sample using the $d_{\nm{uv}}^{\nm{smear}} = 1$ selection are shown by the orange lines, the unconstrained case $P_{\Delta z}$ is shown in green. The NDPV and SDPV values are shown by blue and magenta lines. b) $P_{\Delta z}^{\nm{signal}}$ results corresponding to the $d_{\nm{uv}}^{\nm{smear}} = 1$ selection using all measured $\dz$ values are shown by the orange lines. The results for $\Delta z_{\nm{min}}$ method is shown by the dark red lines. $P_{\Delta z}^{\nm{signal}}$ of the unconstrained case is shown by the green and dark green lines.
}
\label{fig:PDELTAZ10ps}
\end{center}
\end{figure}

In figure~\ref{fig:PDTAG}a) the probability of observing events tagged by ToF detectors on both sides, $P_{\nm{DTAG}}$, is shown as a function of $\mmu$ in the form of colored bands for CDPV (green), NDPV (blue) and SDPV (magenta) samples. Each band represents an envelope of studied edge positions $x_{\nm{min}}^{\nm{det}}$. Top lines correspond to 1~mm, center lines to 1.5~mm and bottom lines to 2~mm and differences with respect to the central values are within 10\%. The effect of using the $\duv$ discriminator cut (realised as $\duvs = 1$) on signal CDPV events is clearly demonstrated by the orange band, to be compared with the green one, corresponding to only the DTAG condition used with no other constraints. Differences between green, magenta and blue bands are easily explained by the fact while the number of pile-up vertices is the same for all samples for a given value of $\mmu$, the number of leading protons is smallest for the NDPV sample and is greater by one for the SDPV and by two for the CDPV sample. That is why the $P_{\nm{DTAG}}$ differences diminish with increasing $\mmu$. Finally, we note that the $P_{\nm{DTAG}}$ values were observed to be insensitive to the time resolution or the hit merging caused by a limited granularity.

The $P_{\rm{DTAG}}$ values resulting from our model and their $\mmu$-dependence can be compared with literature. A combinatorial formula was for instance derived in ref.~\cite{Tasevsky:2014cpa} (see eq.~($1$) there).
As seen in figure~\ref{fig:PDTAG}b), the averaged $P_{\nm{DTAG}}$ probabilities obtained from the toy model as a weighted sum using respective cross-sections of the ND, SD and CD processes, whose sum is denoted as minimum bias (MB), are found consistent with the prediction of the published combinatorial formula for $x_{\nm{min}}^{\nm{det}} = 1.5$~mm. The input parameter to the combinatorial formula is the probability of single-tagging, $A_{ss}$. In our case $A_{\mbox{\tiny ss}}$ is evaluated as

\begin{eqnarray}
A_{\mbox{\tiny ss}} = \frac{\sigma_{\mbox{\tiny SD}}\mbox{A}_{\mbox{\tiny SD}} + \sigma_{\mbox{\tiny CD}} \mbox{A}_{\mbox{\tiny CD}}}{\sigma_{\mbox{\tiny inel.}}}, 
\end{eqnarray}

where the factors 'A' denote the acceptance of FPDs for each process capable of producing leading protons, i.e. fraction of events with $\xi$ values inside the detector acceptance. This in turn means that one can use the analytic prescription to get an idea on how further acceptance changes (e.g. caused by a limited detection efficiency of detectors) propagate to the result. A substitution of $\mmu$ by $\mmu - 1$ is used in the formula and corresponds to the fact that one of the vertices is already occupied by the primary vertex (of a hard-scale event) in the toy model.

\begin{figure}[thh]
\begin{center}
\includegraphics[width=0.47\linewidth, trim={0.5cm 0.2cm 0.5cm 0.5cm},clip]{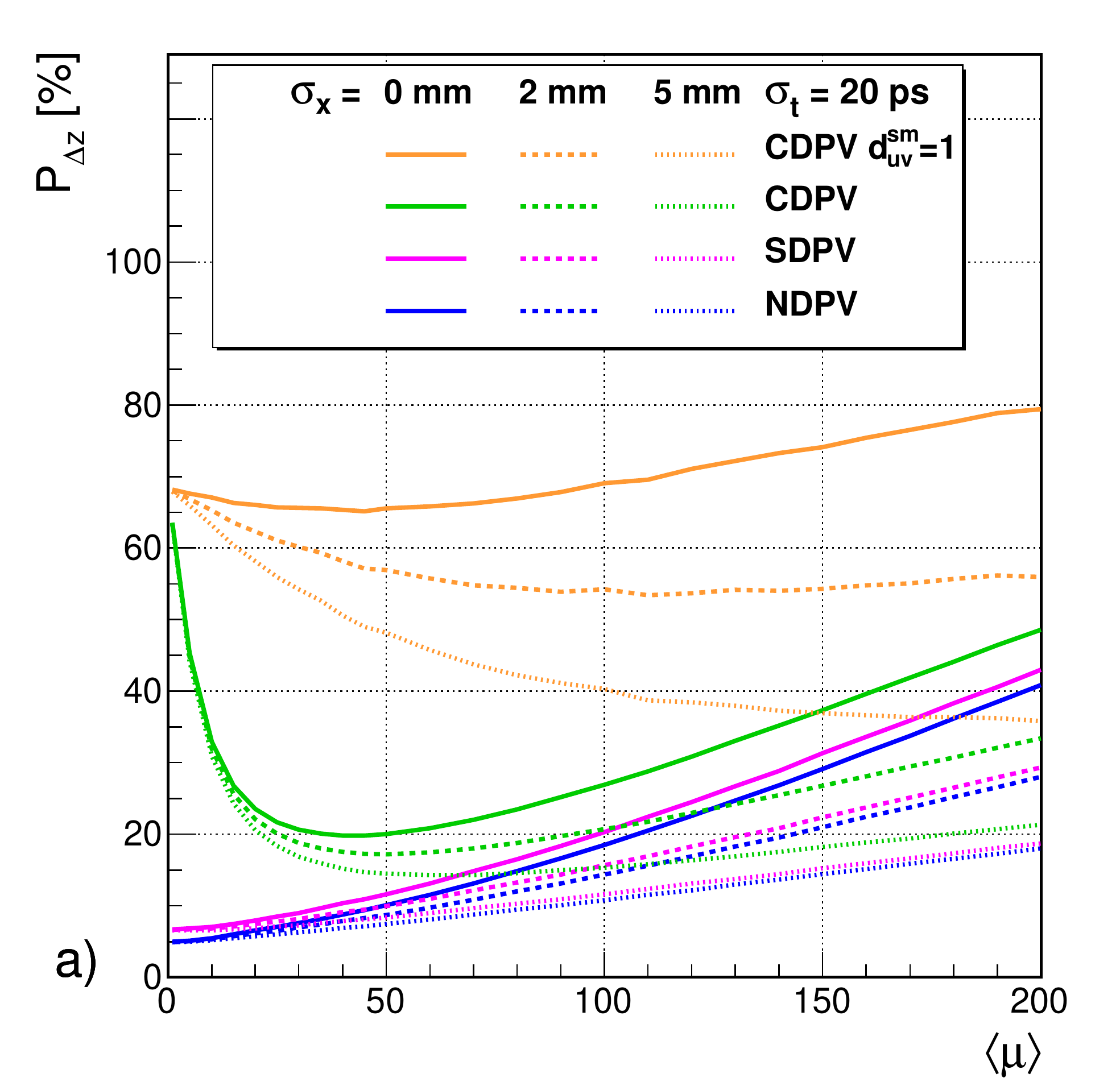}
\includegraphics[width=0.47\linewidth, trim={0.5cm 0.2cm 0.5cm 0.5cm},clip]{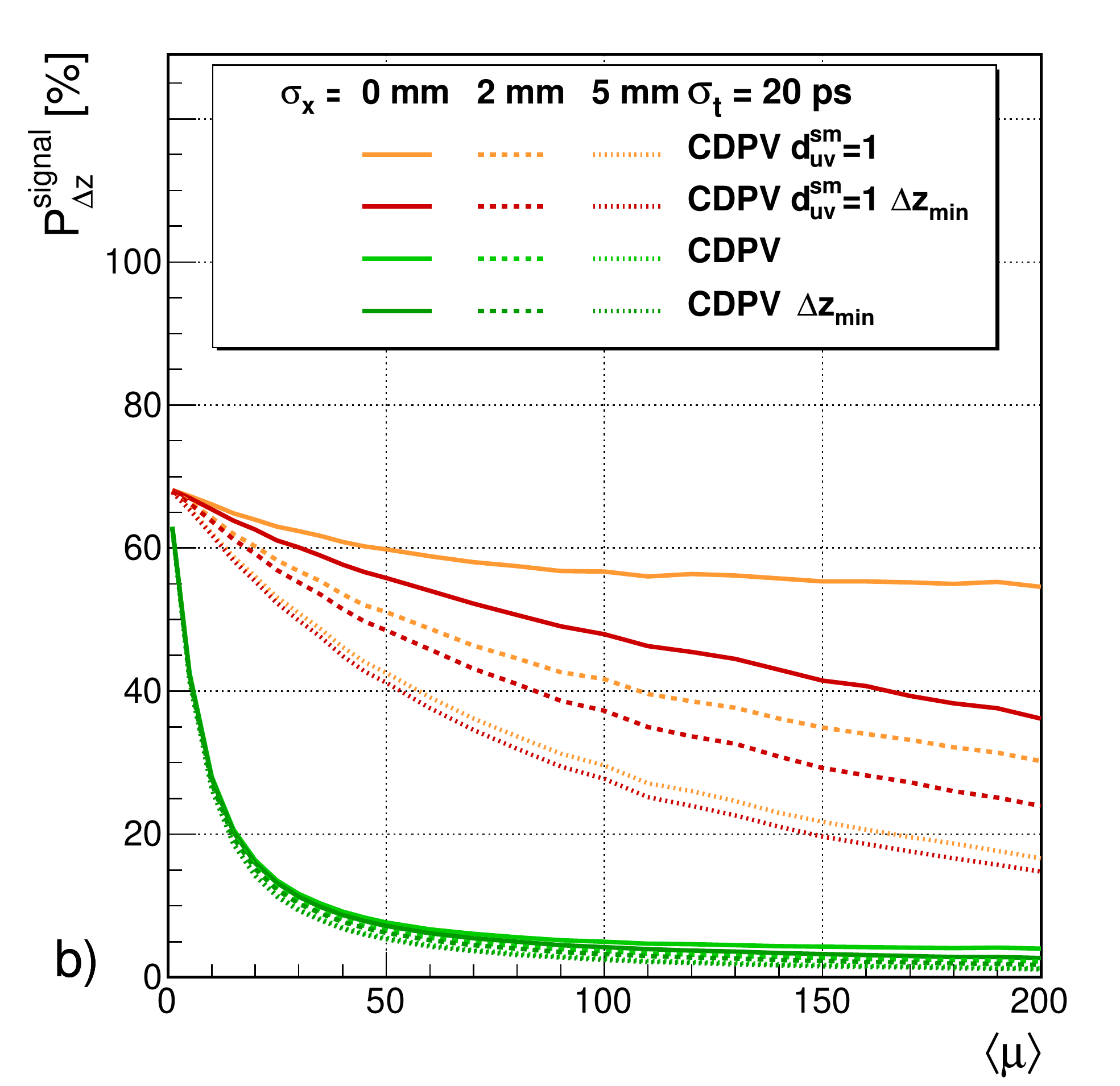}
\caption{The $P_{\dz}$ and $P_{\Delta z}^{\nm{signal}}$ presented in the same manner as in figure~\ref{fig:PDELTAZ10ps} for the timing resolution of $20$~ps.}
\label{fig:PDELTAZ20ps}
\end{center}
\end{figure}

\begin{figure}[bhh]
\begin{center}
\includegraphics[width=0.47\linewidth, trim={0.5cm 0.2cm 0.5cm 0.5cm},clip]{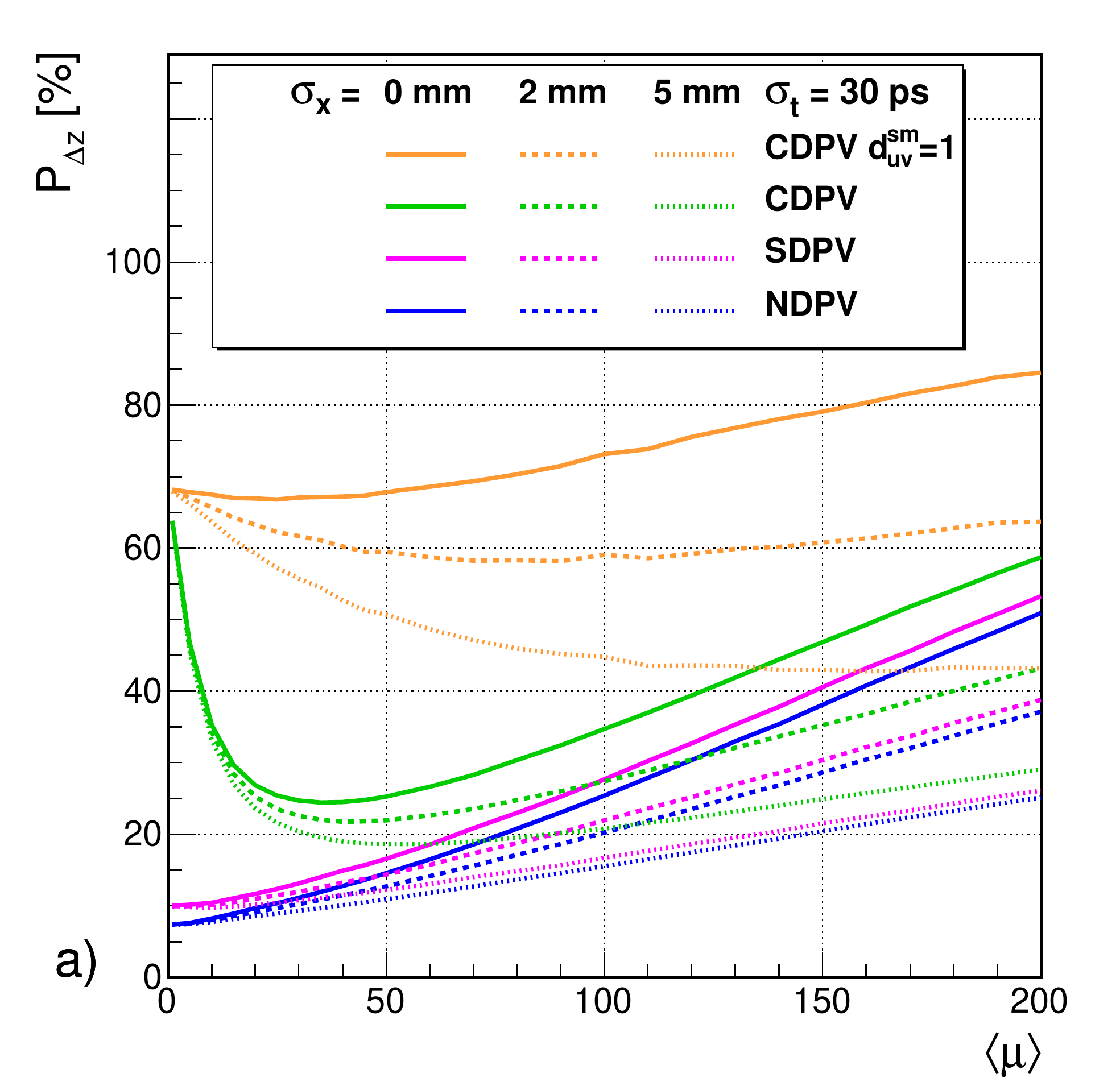}
\includegraphics[width=0.47\linewidth, trim={0.5cm 0.2cm 0.5cm 0.5cm},clip]{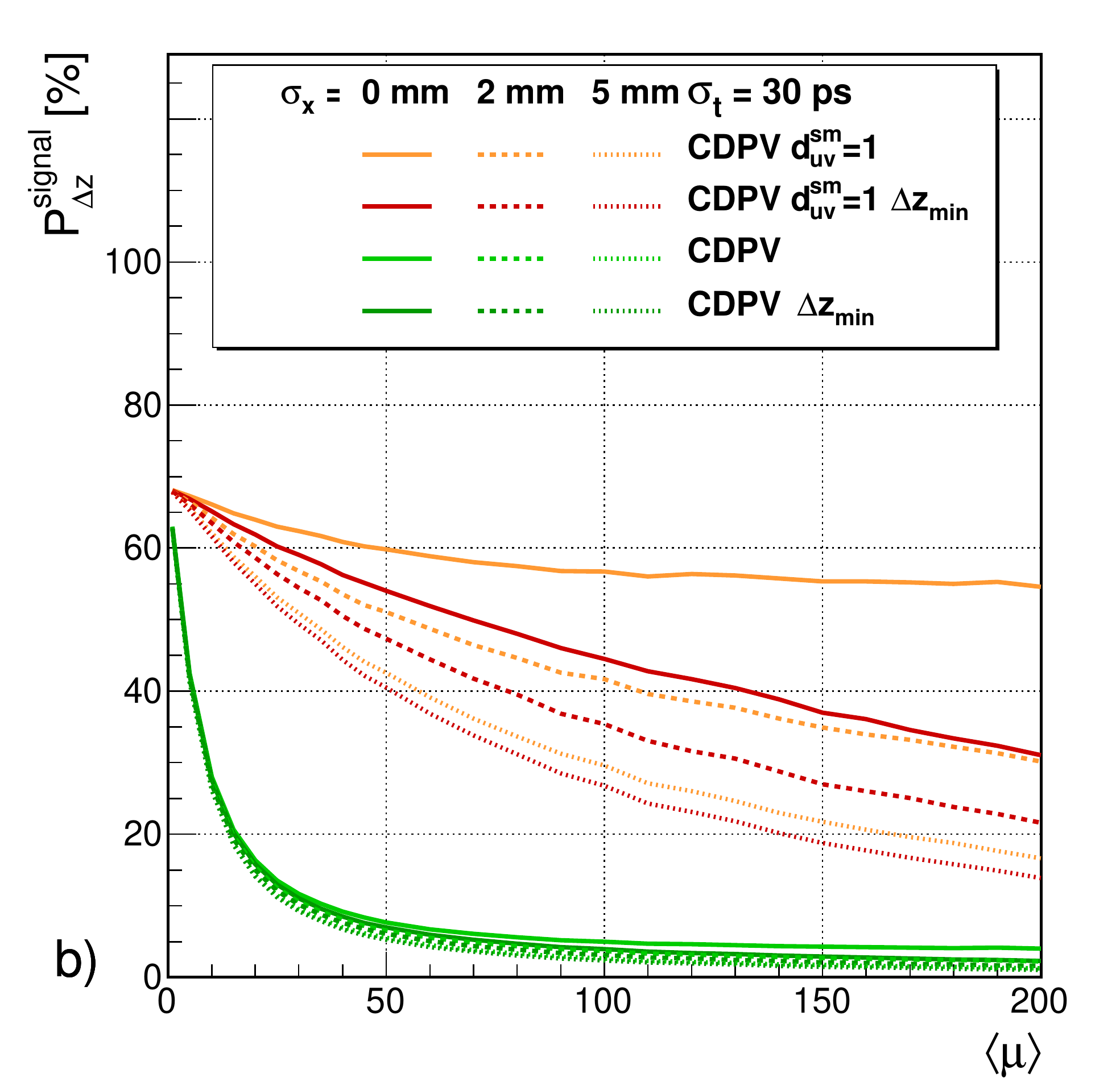}
\caption{The $P_{\dz}$ and $P_{\Delta z}^{\nm{signal}}$ presented in the same manner as in figure~\ref{fig:PDELTAZ10ps} for the timing resolution of $30$~ps.}
\label{fig:PDELTAZ30ps}
\end{center}
\end{figure}

In figures~\ref{fig:PDELTAZ10ps}a),~\ref{fig:PDELTAZ20ps}a) and \ref{fig:PDELTAZ30ps}a)
the $P_{\Delta z}$ probabilities are shown for the same PV samples and $\duv$ cut used as in figure~\ref{fig:PDTAG}, now for three granularity choices of $0, 2$ and $5$~mm in each figure and three timing resolutions of 10, 20 and 30~ps, respectively. Now the different granularities make a difference, the observed probabilities decrease with the granularity parameter $\sigma_{x}$ increasing. The probabilities for CDPV events start at $\sim 68 \%$ and evolve slowly with $\mmu$ for the $\duv$ selected events. The changes in $P_{\dz}$ introduced by changes of granularity do not seem to be dramatic for the event yields. The $P_{\dz}$ values for the CDPV sample with $\duv$ selection relaxed decrease rapidly with $\mmu$ increasing approaching the trends of SDPV and NDPV samples. The fraction of background events passing the $\dz$ cut increases with increasing $\mmu$.

In figures~\ref{fig:PDELTAZ10ps}b),~\ref{fig:PDELTAZ20ps}b) and \ref{fig:PDELTAZ30ps}b)
the $P_{\Delta z}^{\nm{signal}}$ values are shown for the CDPV samples for the granularity and ToF timing resolution as before. The results are presented for events with or without the $\duv$ cut and the $\dz$ and $\dzm$ cut method. The decrease of $P_{\Delta z}^{\nm{signal}}$ selected with $\duv$ cut for all granularities with $\dzm$ method is not surprising, since it only reflects the fact that occasionally the minimum value of $\dz$ in the event can be produced by the pile up interaction. The effect is becoming more visible with timing resolution worsening, because the $\dz$ window widens with $\sigma_{t}$. The observation of low $P_{\dz}^{\nm{signal}}$ values for CDPV events selected using a relaxed $\duv$ cut supports the need for a well designed pre-selections of events by the central detector.

\clearpage
\FloatBarrier


\section{Conclusions}

We have developed a simple model to study the performance of the Time-of-Flight detectors to efficiently
separate central diffractive events from the harsh pile-up environment. The model works with basic assumptions on the transport of leading protons from the interaction point to forward proton detectors
and on time and spatial resolutions of the ToF device. We have provided a generic double-tag probability for the signal and all relevant backgrounds stemming from pile-up interactions, as a function of the time and spatial resolutions of the ToF device and the amount of pile-up per bunch crossing, in the ranges of $\sigma_t$ of $10$--$30$~ps, $\sigma_x$ of $0$--$5$~mm and $\mmu$ of up to $200$. This double-tag probability is to be in an ideal case scaled by selection efficiencies for the signal or rejection efficiencies for backgrounds for each process under study.  

The effect of the time resolution is observed to be rather negligible for the CD signal and more-or-less
linearly increasing with increasing $\sigma_t$ for pile-up backgrounds. 

The effect of the granularity is in general more pronounced for the signal as well as backgrounds and, as expected, while it decreases for the signal, it increases for the backgrounds with increasing $\sigma_x$.
For both these effects, it holds that as the amount of pile-up interactions grows, the effect gets stronger.  

As to the shape of the $\dz = \zpv - \ztpf$ distribution, it was found that the background contribution has two components of different widths -- which may be a useful information for the physics analysis of the real data. 

We have studied two methods of the event selection using the $\dz$ variable, the inclusive and minimum method. While the former is based on looping over all $\dz$ values constructed from all combinations using the list of available $z_{\nm{ToF}}$ values, the latter uses only the $z_{\nm{ToF}}$ value closest to the primary vertex position.
The advantage of the $\dzm$ method is a simpler implementation. Disadvantages are a non-trivial $\dz$ shape and lower probabilities of signal detection in the case of unfavourable time resolutions and granularities of the ToF detectors.

The importance of coupling the ToF detector (or FPD) acceptance with the information from the central detector is demonstrated by the use of $\duv$ discriminator developed specifically for this study.
The derivation of this discriminator is based on a set of generated kinematics ($\mx$ and $\yx$) conveniently transformed to a single-valued observable. The smeared kinematics entering the $\duv$ calculation resulting in a cut on $\duvs$ enhance the fraction of signal events capable of reaching the ToF detectors. In a real data taking the kinematics of the signal are primarily constrained by the trigger followed by another set of cuts relating the central detector observables with those reconstructed from measurements of leading protons in FPDs. Here we only demonstrate that the probability of double-tagging for the central diffractive signal can significantly be enhanced by taking into account kinematical constraints from the central detector.

\section*{Acknowledgements}
The authors gratefully acknowledge the support of the following projects: Karel \v Cern\'y through CZ$.02.1.01/0.0/0.0/16\_019/0000754$ of The Ministry of Education, Youth and Sports the Czech Republic (MEYS); Tom\'a\v s S\'ykora through LM2015056 and LTT17018 of MEYS; Marek Ta\v sevsk\'y through LTT17018 of MEYS; Radek \v Zleb\v c\'ik through Charles University Research Center (UNCE/SCI/013).

\appendix

\section{Double tag discriminator}
\label{app:dbltagdiscriminator}

Here, a discriminator variable, $d_{\nm{uv}}$, of the kinematics $m_{\nm{X}}, y_{\nm{X}}$ generated inside or outside the double tag range is derived. 

From equations~\eqref{eq:mxyx} and \eqref{eq:xiAxiB} it can be seen that $y_{\nm{X}}$ depends linearly on logarithm of $m_{\nm{X}}$ for a fixed value of one of the $\xi$ fractions. For a fixed $\xiB$ one can write

\begin{eqnarray}
y_{\nm{X}}(m_{\nm{X}}, \xiB = \mbox{const}) &=& \frac{1}{2} \ln\frac{\xiA}{\xiB} =   \ln\frac{\sqrt{s\xiA\xiB}}{\sqrt{s\xiB^{2}}} = \nonumber \\
&=& \ln\,m_{\nm{X}} - \ln\,(\sqrt{s}\xiB).
\label{eq:linyxmxdependence}
\end{eqnarray}

The formula~\eqref{eq:linyxmxdependence} corresponds to a linear change of $y_{\nm{X}}$ with $m_{\nm{X}}$ with a positive slope. The negative slope dependence corresponds to a fixed $\xiA$ as

\begin{eqnarray}
y_{\nm{X}}(m_{\nm{X}}, \xiA = \mbox{const}) &=&  -\ln\,m_{\nm{X}} + \ln\,(\sqrt{s}\xiA).
\label{eq:linyxmxdependence2}
\end{eqnarray}

We can also note further symmetry properties defined by four major points in the ($m_{\nm{X}}, y_{\nm{X}}$) kinematic plane (for an idea about symmetries, see figure~\ref{fig:genkin}), i.e.

\begin{alignat}{3}
m_{1} &=&  \sqrt{s} \ximin,\quad y_{1} &=& \frac{1}{2} \mbox{ln}\frac{\ximin}{\ximin} = 0, 
\label{eq:mxyxpoints1} \\
m_{2} &=&  \sqrt{s \ximax \ximin},\quad y_{2} &=& \frac{1}{2} \mbox{ln}\frac{\ximax}{\ximin}\phantom{ = 0},
\label{eq:mxyxpoints2}  \\
m_{3} &=&  \sqrt{s} \ximax,\quad y_{3} &=& \frac{1}{2} \mbox{ln}\frac{\ximax}{\ximax} = 0, \label{eq:mxyxpoints3} \\
m_{4} &=&  \sqrt{s \ximin \ximax},\quad y_{4} &=& \frac{1}{2} \mbox{ln}\frac{\ximin}{\ximax}, \phantom{ = 0} 
\label{eq:mxyxpoints4}
\end{alignat}

where the ($m_{1}, y_{1}$) point denotes the point of mass threshold of double tagging while the ($m_{3}, y_{3}$) indicates the point of maximum achievable mass. The ($m_{2}, y_{2}$) and ($m_{4}, y_{4}$) points represent the points of maximum and minimum rapidity, respectively,  measurable in the double tagged case. If logarithm of a generic base, $\mbox{log}\,m_{\nm{X}}$, is used the $\mbox{log}\,m_{2}$ and $\mbox{log}\,m_{4}$ divide the ($\mbox{log}\,m_{1}$, $\mbox{log}\,m_{3}$) interval in halves since

\begin{eqnarray}
\frac{1}{2}\left( \mbox{log}\,m_{1} + \mbox{log}\,m_{3} \right) & = &\frac{1}{2} \left(\mbox{log}\sqrt{s}\ximin + \mbox{log}\sqrt{s}\ximax \right) = \nonumber \\
&=& \frac{1}{2}\mbox{log}\,s \ximin \ximax = \nonumber\\
&=& \mbox{log}\,\sqrt{s\ximin\ximax} = \mbox{log}\,m_{2} = \mbox{log}\,m_{4}.
\end{eqnarray}

\begin{figure}
\centering
\includegraphics[width=0.99\linewidth,trim={3cm 5cm 4cm 4cm}, clip]{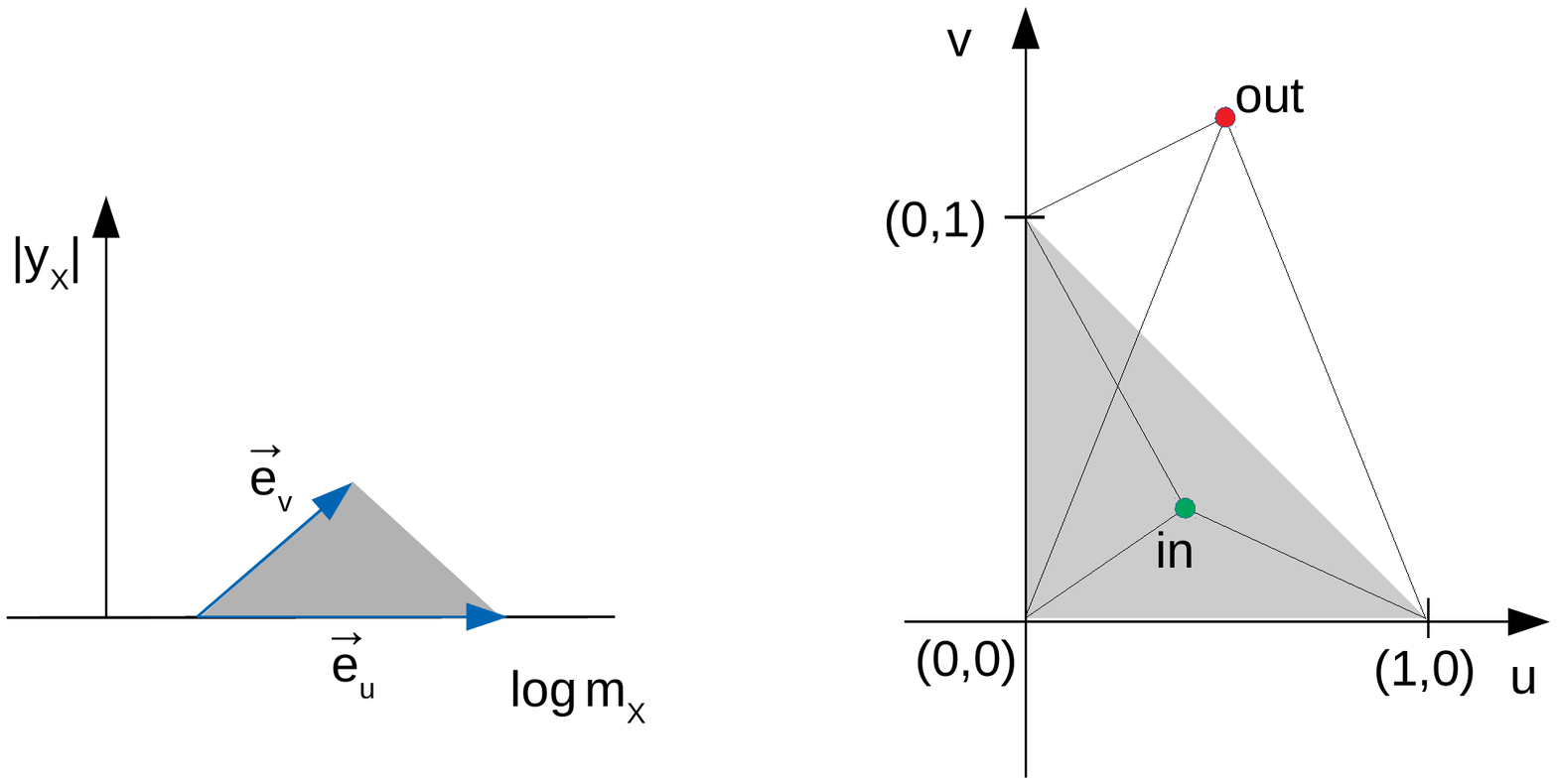}
\caption{The parameterisation of kinematics of double tagged events {in the ($\log\,m_{\nm{X}}$, $|y_{\rm x}|$) plane (left) and in the (u, v) plane (right).}}
\label{fig:uvtriangle}
\end{figure}

The kinematics of the double tag events in the $\log\,m_{\nm{X}}$ and $\left|y_{\nm{X}}\right|$ is represented by a triangle using the $\pm y_{\nm{X}}$ symmetry. If basis vectors defined by two line segments of the kinematic triangle are used ($\vec{e_{u}}, \vec{e_{v}}$), parametric coordinates ($u,v$) define a new parametric triangle ($uv$-triangle), see figure~\ref{fig:uvtriangle}. The points belonging to the triangle satisfy $u,v \in \left<0,1\right>$ and $u + v \leq 1$. The area of the $uv$-triangle is equal to $0.5$. Any point inside a reference convex polygon satisfies the condition that the sum of areas of triangles defined by the sides of the polygon and the tested point equals the polygon area. Any point outside the tested polygon gives a sum larger than the area of the reference polygon. This condition is tested by the calculation of $d_{uv}$ variable defined as

\begin{equation}
 d_{uv} = \frac{0.5}{\sum_{i=1}^{3}A_{i}},
\end{equation}

where the reference value of $0.5$ refers to the $uv$-triangle area and the $A_{i}$ sum term provides a handle on the general point position with respect to the triangle as indicated in figure~\ref{fig:uvtriangle}.

The parametric coordinates in terms of $m_{\nm{X}}$ and $y_{\nm{X}}$ are defined as follows

\begin{eqnarray}
u &=& 
{\frac{1}{\ln\,\frac{m_{X,\nm{max}}}{m_{X,\nm{min}}}}\left[\ln\,\left(\frac{m_{\nm{X}}}{m_{X,min}}\right) - \left|y_{\nm{X}}\right|\right]}, \\
v &=& 
  {\frac{2\left|y_{\nm{X}}\right|}{\ln\,\frac{m_{X,\nm{max}}}{m_{X,\nm{min}}}}} ,
\end{eqnarray}

where the absolute value of $y_{\nm{X}}$ indicates that we employ the $\pm y_{\nm{X}}$ symmetry. The $m_{\nm{X, max}}$ and $m_{\nm{X, min}}$ correspond to $m_{1}$ and $m_{3}$ values defined in eqs.~\eqref{eq:mxyxpoints1} and \eqref{eq:mxyxpoints3}. 

The $\sum_{i = 1}^{3}A_{i}$ term can eventually be written down as

\begin{eqnarray}
\sum_{i = 1}^{3}A_{i} = \frac{1}{2}\left(\left| u\right| + \left| v\right| + \left| u + v - 1 \right|\right),
\end{eqnarray}

leading to the following formula for $d_{uv}$:

\begin{eqnarray}
d_{uv} = \left(\left| u\right| + \left| v\right| + \left| u + v - 1 \right|\right)^{-1}.
\end{eqnarray}

\section{Bunch size propagation to the ToF measurements}
\label{app:widths}
Let us assume that the longitudinal relativistically contracted width of the Gauss-shaped particle bunch in the laboratory frame is $\sigma_{z} = c \sigma_{T}$, where $\sigma_{T}$ is the corresponding 1-$\sigma$ width in time\footnote{The bunch width is often quoted in terms of $4\sigma_{T}$ at the LHC.}. The PDF representing the longitudinal distribution of the beam spot is obtained as a convolution of the PDFs of two bunches moving in time, i.e.
\begin{eqnarray}
f(z) &=& \int_{-\infty}^{+\infty} dt\,\,e^{-\frac{1}{2}\left(\frac{z-ct}{\sigma_{z}}\right)^{2}} e^{-\frac{1}{2}\left(\frac{z+ct}{\sigma_{z}}\right)^{2}} = \nonumber \\
&=& e^{-\frac{1}{2}\left(\frac{z}{\sigma_{z}/\sqrt{2}}\right)^{2}} \underbrace{\int_{-\infty}^{+\infty} dt\,\,e^{-\frac{1}{2}\left(\frac{ct}{\sigma_{z}/\sqrt{2}}\right)^{2}}}_{\nm{constant}}.
\end{eqnarray}

This means that the width along the $z$-direction of the beam spot reads

\begin{equation}
\sigma_{\nm{BS}} = \sigma_{z} / \sqrt{2}.
\label{eq:sigmaBSsigmaZ}
\end{equation}

The $z_{\nm{ToF}}$ vertex position is inferred from the measurements of leading proton arrival times as $z_{\nm{ToF}} = -\frac{c}{2}(t_{\nm{A}} - t_{\nm{B}})$, where $t_{A(B)}$ represents the arrival time of the leading proton to the $A(B)$-side detector. The positive (negative) $z$-side corresponds to the $A$(B)-side. The arrival time is given by the production vertex time, $t_{\nm{pr}}$, retarded (advanced) proportionally to the production vertex $z$-position, $z_{\rm pr}$, as

\begin{equation}
t_{\nm{A(B)}} = \frac{d}{c} -(+) \frac{z_{\nm{pr}}}{c} + t_{\nm{pr}},
\label{eq:tAB}
\end{equation}

where the production vertex values are measured with respect to $t_{\nm{pr}} = 0$ and $z_{\nm{pr}} = 0$ and where equal distances $d$ are assumed from $z_{\nm{pr}} = 0$ to each of the detectors meaning that the distance becomes irrelevant for the $z_{\nm{ToF}}$ calculation.

The values, $z_{\nm{pr}}$ and $ct_{\nm{pr}}$ are distributed with the $\sigma_{BS}$ width. From using the equation~\eqref{eq:tAB} it implies that the width of $ct_{\nm{A(B)}}$ distributions equals $\sqrt{2}\sigma_{\nm{BS}}$. The width of the $z_{\nm{ToF}}$ distribution (given by eq.~\eqref{eq:ztofformula}) obtained from independent production vertices reads

\begin{eqnarray}
\sigma(z_{\nmt{ToF}})^{\nm{indep.}} = \frac{1}{2}\sqrt{\mbox{Var}(z_{\nm{pr}} - ct_{\nm{pr}} + z_{\nm{pr}}^{\prime} + ct_{\nm{pr}}^{\prime})} = \frac{1}{2}\sqrt{4\sigma_{\nm{BS}}^{2}} = \sigma_{\nm{BS}},
\label{eq:sigmazToFunrel}
\end{eqnarray}
where the un-primed (primed) values correspond to the A(B)-side which originate in different unrelated interactions. In the case of interactions leading to the production of two leading protons (central diffraction, CD) registered on both sides, a $\sigma_{\nm{BS}}$ width is expected of the $z_{\nm{ToF}}$ given by

\begin{eqnarray}
\sigma(z_{\nmt{ToF}})^{\nm{CD}} &=& \frac{1}{2}\sqrt{\mbox{Var}(z_{\nm{pr}} - ct_{\nm{pr}} + z_{\nm{pr}} + ct_{\nm{pr}})} = \nonumber \\
&=& \frac{1}{2}\sqrt{\mbox{Var}(2z_{\nm{pr}})} =\frac{1}{2}\sqrt{4\sigma_{\nm{BS}}^{2}} = \sigma_{\nm{BS}}. 
\label{eq:sigmazToFsame}
\end{eqnarray}
    
The distribution of the observable $\Delta z$ defined as $\Delta z = z_{\nm{PV}} - z_{\nm{ToF}}$, where $z_{\nm{PV}}$ is the position of the primary interaction vertex, has two background contributions with different width depending on the type of $z_{\nm{ToF}}$ hypothesis considered. The width of the $\Delta z$ distribution provided by the $z_{\nm{ToF}}$ values obtained from arrival times of unrelated interactions and $z_{\nm{PV}}$ positions of events independent of the ToF tagged ones, i.e. $z_{\nm{PV}}$, $t_{\nm{A}}$ and $t_{\nm{B}}$ independent (called {\it untag} as no leading proton from the primary vertex contributes), is analogously to equation~\eqref{eq:sigmazToFunrel} given by

\begin{eqnarray}
\sigma(\Delta z)^{\nm{untag}} &=& \sqrt{\mbox{Var}\left( z_{\nm{PV}} - \frac{z_{\nm{pr}}}{2} + \frac{ct_{\nm{pr}}}{2} - \frac{z_{\nm{pr}}^{\prime}}{2} - \frac{ct_{\nm{pr}}^{\prime}}{2} \right)} =\nonumber \\
&=& \sqrt{\sigma_{\nm{BS}}^{2} + 4 \frac{\sigma_{\nm{BS}}^{2}}{4}} = \sqrt{2}\sigma_{\nm{BS}}.
\label{eq:sigmadeltazunrelated}
\end{eqnarray}

The same $\Delta z$ width of $\sqrt{2}\sigma_{\nm{BS}}$ is expected in the fraction of $z_{\nm{ToF}}$ values where the arrival times of double tagged CD interactions are measured in events where the $z_{\nm{PV}}$ comes from an independent process, which can be seen from eq.~\eqref{eq:sigmadeltazunrelated} using eq.~\eqref{eq:sigmazToFsame}.

A special case as to the $\Delta z$ width arises when one of the arrival times is actually provided by leading proton originating in the primary interaction process taking place at $z_{\nm{PV}}$ and an unknown time $t_{\nm{PV}}$ (denoted as {\it partial-tag}). With no loss of generality let us assume the A-side values to equal $t_{\nm{pr}} = t_{\nm{PV}}$ and $z_{\nm{pr}} = z_{\nm{PV}}$ in equation~\eqref{eq:sigmadeltazunrelated} leading to

\begin{eqnarray}
\sigma(\Delta z)^{\nm{partial-tag}} &=& \sqrt{\mbox{Var}\left( z_{\nm{PV}} - \frac{z_{\nm{PV}}}{2} + \frac{ct_{\nm{PV}}}{2} - \frac{z_{\nm{pr}}^{\prime}}{2} - \frac{ct_{\nm{pr}}^{\prime}}{2} \right)} =\nonumber \\
&=& \sqrt{\mbox{Var}\left( \frac{z_{\nm{PV}}}{2} + \frac{ct_{\nm{PV}}}{2} - \frac{ct_{\nm{pr}}^{\prime}}{2} - \frac{z_{\nm{pr}}^{\prime}}{2} \right)} =
\nonumber \\
&=&
\sqrt{4\frac{\sigma_{\nm{BS}}^{2}}{4}} = \sigma_{\nm{BS}}.
\label{eq:sigmadeltazpartialtag}
\end{eqnarray}

\bibliographystyle{JHEP_Karel}

\bibliography{pileupstudy}

\end{document}